\documentclass[aps,prc,floatfix,twocolumn,superscriptaddress]{revtex4-1}

\usepackage[]{graphicx}
\usepackage{color}

\usepackage{amsmath,amssymb,amsfonts}

\usepackage{slashed}

\begin{document}

\title{Charmed hadron chemistry and flow in heavy and light ion collisions at the LHC}

\author{Yu-Fei Liu}
\affiliation{Institute of Particle Physics and Key Laboratory of Quark and Lepton Physics (MOE), Central China Normal University, Wuhan, Hubei, 430079, China}

\author{Wen-Jing Xing}
\email{wenjing.xing@mails.ccnu.edu.cn}
\affiliation{Institute of Frontier and Interdisciplinary Science, Shandong University, Qingdao, Shandong 266237, China}

\author{Xiang-Yu Wu}
\affiliation{Institute of Particle Physics and Key Laboratory of Quark and Lepton Physics (MOE), Central China Normal University, Wuhan, Hubei, 430079, China}
\affiliation{Department of Physics, McGill University, Montreal, Quebec, H3A 2T8, Canada}

\author{Shanshan Cao}
\affiliation{Institute of Frontier and Interdisciplinary Science, Shandong University, Qingdao, Shandong 266237, China}

\author{Guang-You Qin}
\email{guangyou.qin@ccnu.edu.cn}
\affiliation{Institute of Particle Physics and Key Laboratory of Quark and Lepton Physics (MOE), Central China Normal University, Wuhan, Hubei, 430079, China}

\author{Shu-Qing Li}
\email{lisq79@jnxy.edu.cn}
\affiliation{School of Physical Science and Intelligent Engineering, Jining University, Qufu, Shandong, 273155, China}

\date{\today}
\begin{abstract}

We study the charmed meson and baryon production and elliptic flow in ultra-relativistic nucleus-nucleus collisions at the LHC energies.
The space-time evolution of quark-gluon plasma (QGP) produced in these energetic collisions is obtained via the (3+1)-dimensional CLVisc hydrodynamics model, the heavy quark dynamics inside the QGP is simulated using an improved Langevin model that incorporates both elastic and inelastic parton energy loss processes, and the heavy quark hadronization is simulated utilizing a comprehensive coalescence-fragmentation model.
Using our combined approach,
we first calculate charmed hadron ratios, $\Lambda_c/D^0$ and $D_s/D^0$, as well as their elliptic flow ($v_2$) as a function of transverse momentum ($p_T$) for different centralities in Pb+Pb collisions at $\sqrt{s_{NN}}=5.02$~TeV.
Due to strangeness enhancement and parton coalescence effects, $D_s/D^0$ and $\Lambda_c/D^0$ ratios increase from peripheral to central collisions, and such centrality dependence for $\Lambda_c/D^0$ is stronger than $D_s/D^0$.
We further predict the $p_T$ and centrality dependences of charmed hadron chemistry and $v_2$ in smaller Xe+Xe, Ar+Ar and O+O collisions at the LHC energies.
Strong centrality and system size dependences for $\Lambda_c/D^0$ and $D_s/D^0$ ratios are observed across four collision systems.
As for charmed hadron flow, both system size and collision geometry are important to understand the centrality dependence of $v_2$ in different collision systems.
Our study provides a significant reference for studying heavy quark evolution and hadronizaiton in large and small systems in relativistic nuclear collisions.

\end{abstract}
\maketitle

\section{Introduction}

High-energy nucleus-nucleus collisions performed at the Relativistic Heavy-Ion Collider (RHIC) and the Large Hadron Collider (LHC) have created the hot and dense quark-gluon plasma (QGP), a new state of matter consisting of color-deconfined quarks and gluons, as predicted by Quantum Chromodynamics (QCD).
Anisotropic flow and jet quenching have been regarded as two most important signatures for the formation of QGP in these energetic nuclear collisions.
On one hand, the strong collective motion of the excited QCD matter and the azimuthal anisotropy of low transverse momentum hadrons emitted from the QGP can be successfully explained by relativistic hydrodynamic simulation with extremely small specific shear viscosity~\cite{Adams:2003zg, Aamodt:2010pa, ATLAS:2011ah, Romatschke:2017ejr, Rischke:1995ir, Heinz:2013th, Gale:2013da, Huovinen:2013wma}, suggesting that the QGP produced in relativistic heavy-ion collisions is strongly interacting and behaves like a perfect fluid.
On the other hand, the suppression of high transverse momentum hadrons and reconstructed jets in nucleus-nucleus collisions compared to proton-proton collisions scaled by the number of binary collisions, and other related phenomena, can be well described by jet-medium interaction and jet quenching effects~\cite{Wang:1991xy, Gyulassy:2003mc, Majumder:2010qh, Qin:2015srf, Blaizot:2015lma, Cao:2020wlm, Cao:2022odi}.
Sophisticated studies have shown that the values of jet quenching parameter $\hat{q}$ in the hot QGP is about two orders of magnitude larger that in cold nuclear matter, indicating the deconfined quark and gluon degrees of freedom inside the excited nuclear matter~\cite{Bass:2008ch,Burke:2013yra,Cao:2021keo}.

Among various probes of the QGP medium, heavy (charm and bottom) quarks have been considered to be the unique probes due to their masses much larger than $\Lambda_{QCD}$ and the temperature of QGP~\cite{Dong:2019byy,Andronic:2015wma,He:2022ywp}. They are mostly produced from initial hard partonic collisions in the early stage of heavy-ion collisions, thus have the potential to probe the entire evolution history of the expanding QGP. Their production cross sections can also be calculated within perturbative QCD.
Meanwhile, heavy quarks also provide versatile tools to probe the QGP properties due to their comprehensive coverage in $p_T$.
The low $p_T$ heavy quarks can be used to study the diffusion, flow and thermalization of heavy quarks inside the QGP, which offers an excellent tool to probe the non-perturbative regime of the QCD matter~\cite{Moore:2004tg,Xing:2021xwc,Liu:2023rfi}.
At very high $p_T$, heavy quarks are ideal probes to study the interplay between collisional and radiative parton energy loss mechanisms in the QGP and the flavor dependence of jet quenching in heavy-ion collisions~\cite{Djordjevic:2013pba,Xing:2019xae}.
At intermediate $p_T$, heavy quarks provide a powerful tool to study the hadronization process of partons~\cite{Plumari:2017ntm,He:2019vgs,Cho:2019lxb,Cao:2019iqs}.
In the past decades, tremendous studies have been performed on understanding the production, evolution and interaction of heavy quarks inside the cold and hot nuclear medium, and the hadronization of heavy quarks in the final state of relativistic nuclear collisions
~\cite{Gossiaux:2006yu,Qin:2009gw,Das:2010tj,Uphoff:2011ad,He:2011qa,Young:2011ug,Alberico:2011zy,Fochler:2013epa,Nahrgang:2013saa,Cao:2013ita,Djordjevic:2013xoa,Cao:2015hia,Das:2015ana,Song:2015ykw,Cao:2016gvr,Kang:2016ofv,Prado:2016szr,Cao:2017crw,Xu:2017obm,Liu:2017qah,Rapp:2018qla,Cao:2018ews,Li:2018izm,Ke:2018tsh,Li:2019wri,Katz:2019fkc,Li:2020kax,Chen:2021uar,Liu:2021dpm,Yang:2023rgb}.
There have been a wealthy of experimental data on the nuclear modification factor, collective flow, and particle ratios of heavy flavor hadrons at RHIC and the LHC~\cite{PHENIX:2006iih,STAR:2014wif,STAR:2017kkh,CMS:2017qjw,CMS:2017vhp,ALICE:2017pbx,ALICE:2018lyv,STAR:2018zdy,Acharya:2018ckj,STAR:2019clv,Adam:2019hpq,Vermunt:2019ecg,ALICE:2020iug,ATLAS:2020yxw,ALICE:2021kfc,ALICE:2021bib}.

In this work, we focus on the charmed meson and baryon production and elliptic flow in Pb+Pb, Xe+Xe, Ar+Ar and O+O collisions at the LHC energies.
Our study of both heavy and light ion collisions is motivated by the strong interest in understanding the evolution dynamics of large and small collision systems by scanning collision systems with different sizes.
In particular, the LHC has proposed to run O+O collisions in the near future.
In order to perform such systematic study, we use (3+1)-dimensional CLvisc hydrodynamics model~\cite{Pang:2009zm, Pang:2018zzo, Wu:2021fjf} to simulate the space-time evolution of QGP produced in both heavy and light ion collisions.
The dynamical evolution of heavy quarks inside the QGP is simulated using our improved Langevin model~\cite{Cao:2013ita, Cao:2015hia, Li:2019wri} that includes both collisional and radiative interactions between heavy quarks and the QCD medium.
The hadronization of heavy quarks after escaping from the QGP is calculated using an advanced coalescence-fragmentation model~\cite{Cao:2019iqs}.
Using our combined model,
we first calculate the yields and elliptic flow of charmed hadrons $D^0$, $D_s$ and $\Lambda_c$ for different centralities in Pb+Pb collisions at $\sqrt{s_{NN}}=5.02$~TeV.
We find that $D_s/D^0$ and $\Lambda_c/D^0$ ratios increase from peripheral to central collisions due to the strangeness enhancement and parton coalescence effects, and the centrality dependence is stronger for $\Lambda_c/D^0$ than $D_s/D^0$ because of the baryon-to-meson enhancement effect.
To study the system size dependence across different collision systems, we perform the calculations for the $p_T$ and centrality dependences of the $\Lambda_c/D^0$ and $D_s/D^0$ ratios and $v_2$ of $D^0$, $D_s$ and $\Lambda_c$ in Xe+Xe, Ar+Ar and O+O collisions at the LHC energies.

The paper is organized as follows. In the next section, we present the setups for hydrodynamics simulation of QGP evolution in Pb+Pb, Xe+Xe, Ar+Ar and O+O collisions at the LHC energies.
In Sec. III, we present the details of our Langevin approach on how to simulate heavy quark evolution inside the QGP.
Sec. IV describes the hybrid coalescence-fragmentation formalism for heavy quark hadronization.
In Sec. V, we present the numerical results for charmed hadron ratios ($\Lambda_c/D^0$ and $D_s/D^0$), and the elliptic flow $v_2$ of $D^0$, $D_s$ and $\Lambda_c$ in Pb+Pb, Xe+Xe, Ar+Ar and O+O collisions.
Sec. VI contains our summary.

\section{Langevin approach for heavy quark evolution in QGP}

In this work, we use our modified Langevin approach~\cite{Cao:2013ita} to simulate the evolution of heavy quarks inside the QGP medium.
It includes both quasi-elastic scattering and medium-induced gluon bremsstrahlung processes as follows~\cite{Cao:2013ita}:
 \begin{align}
  \label{eq:Langevin}
  \frac{d\vec{p}}{dt} = -\eta _{D}(p)\vec{p}+\vec{\xi}+\vec{f_{g}}.
 \end{align}
In the right hand side of the above equation, the first and the second terms denote the drag force and the thermal random force experienced by a heavy quark when it interacts with the QCD medium constituents.
Typically, one assumes that the thermal force $\vec{\xi}$ is independent of heavy quark momentum and satisfies the following correlation function of white noise:
\begin{align}
\langle\xi^{i}(t)\xi^{j}(t^{\prime})\rangle=\kappa\delta^{ij}\delta(t-t^{\prime}),
\end{align}
where $\kappa$ is the momentum diffusion coefficient of heavy quarks and determines the interaction strength of the thermal force.
The drag coefficient $\eta_{D}$ can be determined from $\kappa$ using the fluctuation-dissipation theorem: $\eta_{D}(p)=\kappa/(2TE)$.
Then the spatial diffusion coefficient $D_{s}$ can be related to $\kappa$ as: $D_{s}\equiv T/[M\eta_{D}(0)]=2T^{2}/\kappa$.
To include the effects of medium-induced gluon radiation process, Ref.~\cite{Cao:2013ita} has introduced a third term $\vec{f}_g$ to the above Langevin equation.
This new term describes the recoil force $\vec{f_{g}}=-d\vec{p}_{g}/dt$ experienced by the heavy quark when it emits a gluon with momentum $\vec{p}_g$ induced by the multiple scatterings with the medium constituents.

To simulate the medium-induced gluon radiation process, we compute the probability of medium-induced gluon radiation $P_\mathrm{rad}(t,\Delta t)$ during a time interval $(t,t+\Delta t)$ via the following formula,
 \begin{align}
 \label{eq:gluonProb}
 P_\mathrm{rad}(t,\Delta t) = \langle N_{g}(t,\Delta t)\rangle = \Delta t\int dxdk_{\perp}^{2}\frac{dN_{g}}{dxdk_{\perp}^{2}dt},
 \end{align}
where $\langle N_{g}(t,\Delta t)\rangle$ represents the average number of emitted gluons during this time interval; $x$ and $k_\perp$ are the energy fraction and transverse momentum of the emitted gluon with respect to the parent heavy quark.
Here, we choose sufficiently small values for the time interval $\Delta t$ in order to guarantee $\langle N_{g}(t,\Delta t)\rangle <1$.
Then the average number $\langle N_{g}(t,\Delta t)\rangle$ can be interpreted as probability.

In our study, we take higher-twist (HT) energy loss formalism~\cite{Guo:2000nz,Majumder:2009ge,Zhang:2003wk,Zhang:2018nie} to study the medium-induced radiative process. In HT formalism, the spectrum of medium-induced gluon radiation off a heavy quark reads as:
\begin{align}
\label{eq:gluonSpectrum}
\frac{dN_{g}}{dxdk_{\perp}^{2}dt}=\frac{2\alpha_{s}P(x)k_\perp^4 \hat q}{\pi ({k_{\perp}^{2}+x^{2}M^{2}})^{4}}\sin^{2}\left(\frac{t-t_{i}}{2\tau _{f}}\right).
\end{align}
In the above formula, $\alpha _s$ is the strong coupling which runs with the exchanged transverse momentum squared $k_\perp^2$ and $P(x)$ is the parton splitting function for $Q\rightarrow Qg$ process. The transport coefficient parameter $\hat q$ is defined as the transverse momentum squared exchanged with the medium per unit time. It controls the amount of medium-induced gluon radiation and relates to the heavy quark diffusion coefficient via $\hat q = 2\kappa C_{A}/C_{F}$, where $C_A$ and $C_F$ are color factors of gluon and quark, respectively. $t_i$ is the starting time of the radiation process, and $\tau _{f}=2Ex(1-x)/(k_{\perp}^{2}+x^{2}M^{2})$ represents the formation time of the gluon radiation with energy $xE$ and transverse momentum $k_\perp$, where $E$ and $M$ are the energy and mass of heavy quarks.
When using the above formula to sample the energy and transverse momentum of the medium-induced gluons, we impose a lower cutoff for the gluon energy $E_g = xE > \pi T$. Such cutoff mimics the balance between the gluon emission and absorption processes around the thermal scale, as we have not included the absorption process in the current implementation.

With the above setup, our improved Langevin approach contains only one free parameter, which we choose to be the dimensionless parameter $D_s (2\pi T)$. It quantifies the interaction strength between heavy quarks and the QGP medium. In this study, we take $D_\mathrm{s}(2\pi T)=4$ for all four collision systems. Such value have been show to provide a reasonable description of heavy flavor meson observables in Pb+Pb collisions at the LHC \cite{Li:2020kax,Li:2021xbd}.

To simulate heavy quark evolution through QGP, we need the initial conditions for heavy quark production.
The spatial distribution of the initial heavy quarks is sampled according to the binary collision vertices as calculated from Monte-Carlo Glauber model~\cite{Miller:2007ri}.
The momentum distribution of the initial heavy quarks is calculated using the fixed-order-next-to-leading-log (FONLL) calculation~\cite{Cacciari:2001td,Cacciari:2012ny,Cacciari:2015fta} convoluted with the CT14NLO~\cite{Dulat:2015mca} parton distribution function.
After their initial production, the subsequent evolution of heavy quarks inside the QGP medium is then simulated via the above Langevin model.
The space-time profile of the expanding QGP medium is computed using the CLVisc hydrodynamic model, which will be describe in the next section.

\section{Hydrodynamics simulation of QGP evolution}

In this study, we utilize the (3+1)-dimensional viscous hydrodynamics model (CLVisc) \cite{Pang:2009zm, Pang:2018zzo, Wu:2021fjf} to simulate the dynamical evolution of QGP produced in Pb+Pb, Xe+Xe, Ar+Ar and O+O collisions at the LHC energies.
The initial three-dimensional entropy density distribution $S(\tau_0, x, y, \eta_s)$ of the QGP is obtained from the entropy density distribution $s(x,y)$ in the transverse plane and the longitudinal envelop function $H(\eta)$ as follows:
\begin{align}
S(\tau_0, x,y,\eta_s) = Ks(x,y)H(\eta_s),
\end{align}
where $K$ is a normalization factor.
For Pb+Pb and Xe+Xe collisions at the LHC, $K$ factors are determined by comparing to the experimental data on final charged hadron spectra in the most central collisions~\cite{Adam:2016ddh, Acharya:2018hhy}.
As for Ar+Ar and O+O collisions at the LHC energies, the following empirical power-law formula~\cite{Acharya:2018hhy} is used to calculate the final charged hadron spectra:
\begin{align}
\frac{2}{\left< N_{\text{part}} \right>} \left.\left<  \frac{dN_{\text{ch}}}{d\eta}\right>\right|_{|\eta|<0.5}=0.7455 \left(\frac{s_{\rm NN}}{{\rm GeV}^2}\right)^{0.1538}.
\end{align}
The above formula is obtained by fitting to the multiplicity data of Au+Au collisions at RHIC and Pb+Pb collisions at the LHC at various colliding energies.
For the entropy density distribution $s(x,y)$ in the transverse plane, the TRENTo model~\cite{Moreland:2014oya} is used.
In this model, the positions of nucleons inside lead, zenon, argon and oxygen nuclei are sampled according to the Woods-Saxon distribution,
\begin{align}
\rho(r,\theta) = \rho_0 \frac{1+w\frac{r^2}{R^2(\theta)}}{1+\exp\left(\frac{r-R(\theta)}{d}\right)},
\end{align}
where $\rho_0$ denotes the average nuclear density, $d$ is a diffusion parameter at the surface, and $R(\theta)$ is the radius parameter.
For the nuclei studied in this work, $R(\theta)$ is given by:
\begin{align}
R(\theta)=R_0[1+\beta_2 Y_{20}(\theta) + \beta_4 Y_{40}(\theta)],
\end{align}
where $Y_{nl}(\theta)$ is the spherical harmonic functions, $w$, $\theta_2$ and $\beta_4$ are the parameters that control the deviations of nucleus distribution from the spherical shape.
Table.~\ref{tab:Woods-Saxon} shows the values of various parameters used in Woods-Saxon distribution and the TRENTo model for Pb+Pb, Xe+Xe, Ar+Ar and O+O collisions at the LHC energies.

\begin{table}[h]
\centering
\begin{tabular}{|c|c|c|c|c|c|c|c|}
\hline
	System &  $R_0$ (fm) & $d$ (fm) & $w$ & $\beta_2$ & $\beta_4$ & $\sigma_{\rm inel}^{\rm NN}$(mb)\\ \hline
	PbPb @ 5.02TeV & 6.62 & 0.546 & 0     & 0 & 0 & 70 \\ \hline
    XeXe @ 5.44TeV & 5.4 & 0.59 & 0     & 0.18 & 0 & 70 \\ \hline
	ArAr @ 5.85TeV & 3.53 & 0.542 & 0      & 0 & 0 & 71 \\ \hline
	OO @ 6.5TeV & 2.608 & 0.513 & -0.051 & 0 & 0 & 72.5 \\ \hline
\end{tabular}
\caption{The values of the parameters used in the Woods-Saxon distribution and TRENTo models for different collision systems~\cite{Acharya:2018hhy,Sievert:2019zjr}.}
\label{tab:Woods-Saxon}
\end{table}

In the TRENTo model, the determination of participant nucleons relies on the effective patron cross section, denoted as $\sigma_g$, which is tuned to fit the inelastic nucleon-nucleon cross sections $\sigma_{\rm inel}^{\rm NN}$. Then the thickness function $T_A(x,y)$ and $T_B(x,y)$ can be calculated by employing Gaussian smearing from each participant nucleons, along with a random gamma distribution weight. As a result, the reduced entropy density distribution $s(x,y)$ in the transverse plane is obtained via performing a the generalized average from the thickness functions of two colliding nuclei:
\begin{align}
s(x,y) = \left[\frac{T_A^p(x,y) + T_B^p(x,y)}{2}\right]^{\frac{1}{p}},
\end{align}
where $p$ is a parameter in the model, which is set as $p=0$ in our work, i.e., $s(x,y) = \sqrt{T_A(x,y) T_B(x,y)}$.
The longitudinal envelop function $H(\eta_s)$ is parametrized as follows~\cite{Pang:2018zzo},
\begin{align}
H(\eta_s) = \exp\left[ -\frac{(|\eta_s| - \eta_0)^2}{2\sigma^2_{\eta_s}}\theta(|\eta_s| - \eta_0) \right],
\end{align}
where the parameters $\eta_0$, $\sigma_{\eta_s}$ should be fitted to the pseudorapidity distribution of final particle yields.
Table~\ref{tab:hydro} shows the values of the overall normalization $K$, the initial time $\tau_0$ and the longitudinal parameters $\eta_0$ and $\sigma_{\eta_s}$ used for the initial conditions of the hydrodynamical simulation of the bulk matter produced in Pb+Pb, Xe+Xe, Ar+Ar and O+O collisions at the LHC energies.
In this work, we use smooth initial conditions of entropy density distribution for hydrodynamic simulation by averaging over 5000 Trento events in each centrality class.

\begin{table}[h]
\centering
\begin{tabular}{|c|c|c|c|c|c|c|}
\hline
	System  & $K$ & $\tau_0$(fm) & $\eta_0$ & $\sigma_{\eta_s}$\\ \hline
	PbPb @ 5.02TeV   & 154  & 0.6 & 1.7 & 2.0\\ \hline
	XeXe @ 5.44TeV  & 155  & 0.6 & 1.8 & 2.23 \\ \hline
	ArAr @ 5.85TeV   & 160  & 0.6 & 1.7 & 2.0\\ \hline
	OO @ 6.5TeV     & 180  & 0.6 & 1.7 & 2.0 \\ \hline
\end{tabular}
\caption{The values of parameters for hydrodynamic simulation of the bulk QGP matter in different collision systems.}
\label{tab:hydro}
\end{table}

In the (3+1)-dimensional CLVisc hydrodynamics model~\cite{Pang:2009zm, Pang:2018zzo, Wu:2021fjf}, we solve the following equations of motion for the energy-momentum tensor $T^{\mu\nu}$ and the shear stress tensor $\pi^{\mu\nu}$,
\begin{eqnarray}
& &\partial_{\mu}T^{\mu\nu}  = 0, \\
& &\pi^{\mu\nu} = \eta_v \sigma^{\mu\nu} - \tau_{\pi}\left[\Delta^{\mu\nu}_{\alpha \beta}u^{\lambda}\partial_{\lambda}\pi^{\alpha \beta} + \frac{4}{3}\pi^{\mu\nu}\theta \right],
\end{eqnarray}
where $\sigma^{\mu\nu}$ is the symmetric shear tensor, and $\theta$ is the expansion rate.
In this study, the above equations are solved in Milne coordinate using the Kurganov-Tadmor (KT) algorithm.
The specific shear viscosity is set as $\eta_v/s = 0.16$ and the relaxation time is set as $\tau_{\pi} = {3\eta_v}/{(sT)}$.
In order to close the system, the equations of state of the QCD matter is taken from the {\rm s95-pce-165} parametrization~\cite{Huovinen:2009yb}.
The transition from the hydrodynamics evolution to hadrons is performed via the Cooper-Frye formula at the switching temperature $T_{\rm sw} = 137$~MeV.
Using the above setups, we perform the hydrodynamic calculation for Pb+Pb, Xe+Xe, Ar+Ar and O+O collisions at the LHC energies.
Our calculation can give reasonable description of the soft hadron spectra in Pb+Pb collisions at $\sqrt{s_{NN}}=5.02$~TeV and Xe+Xe collisions at $\sqrt{s_{NN}}=5.44$~TeV.
Our hydrodynamics calculation also provide the predictions for the QGP profiles for Ar+Ar and O+O collisions used in this study.

\section{Hadronization for heavy quark hadronization}

When the local temperature of the medium drops below the transition temperature $T_c = 160$~MeV, the interaction betwen heavy quarks and the medium ceases and heavy quarks are then converted to heavy flavor hadrons.  In this work, we utilize a comprehensive coalescence-fragmentation model developed in Ref.~\cite{Cao:2019iqs} for heavy quark hadronization.
In this hybrid hadronization model, the probability for a heavy quark to form a heavy flavor hadron by coalescence with the light quarks in the QGP medium is calculated from the wave function overlap between free quark states and hadronic bound states.
In our study, we include both $s$ and $p$-wave states for heavy flavor hadrons.
After obtaining the probability, we first determine whether a heavy quark forms heavy flavor hadron via the coalescence mechanism. If it does not hadronize through the coalescence with the light quarks from QGP, we then put it into Pythia~\cite{Sjostrand:2006za} simulation and let it fragment into hadrons.
For analyzing the final state heavy flavor observables (e.g., particle yield and elliptic flow), we use heavy flavor hadrons produced from both coalescence and fragmentation processes.

In the quark coalescence model, one calculates the momentum distribution of produced hadrons via the following formula,
\begin{align}
\label{eq:distributionH}
f_h(\bm{p}_h)=\int \Big[\prod_{i} d\bm{p}_i f_i (\bm{p}_i)\Big] W(\{\bm{p}_i\}, \bm{p}_h)\delta (\bm{p}_h-\sum_{i} \bm{p}_i),
\end{align}
where $\bm{p}_h$ is the momentum of the hadron, $\bm{p}_i$ is the momentum of the constituent quark, and $W(\{\bm{p}_i\}, \bm{p}_h)$ is the Wigner function describing the coalescence probability for free quark states to form hadron bound states.

Taking the meson formation as an example. If one assumes a harmonic oscillator potential for the quark-antiquark system, the Wigner functions for $s$ and $p$-wave mesons can be obtained as follows~\cite{Baltz:1995tv,Chen:2006vc}:
\begin{align}
\label{eq:WignerS}
&W_s=g_h\frac{(2\sqrt{\pi}\sigma)^3}{V} e^{-\sigma^2\bm{k}^2},\\
\label{eq:WignerP}
&W_p=g_h\frac{(2\sqrt{\pi}\sigma)^3}{V} \frac{2}{3}\sigma^2\bm{k}^2 e^{-\sigma^2\bm{k}^2}.
\end{align}
In the above equation, the Wigner functions have been averaged over the volume $V$; the degeneracy factor $g_h$ includes the spin and color degrees of freedom;
$\bm{k}$ is the relative momentum between the two constituent quarks in the the rest frame of the meson; $\sigma$ is the width of the relative momentum distribution, which can be related to the the frequency of $\omega$ the harmonic oscillator as $\sigma=1/\sqrt{\mu\omega}$, with $\mu$ being the reduced mass of two constituent quark system.
The extension of the coalescence model to a three constituent quark system to form a baryon is relatively straightforward. One first combines two quarks and then combines their center-of-momentum with the third quark. Note that for the formation of baryons, all possible inner configurations (e.g., which two quarks are combined first) should be symmetrized.

In this work, the charm quark mass is taken as $m_c=1.8$~GeV and the thermal masses of light quarks are taken as $m_{u}=m_{d}=0.3$~GeV and $m_s=0.4$~GeV in the coalescence model.
We include all possible $s$ and $p$-wave hadron states allowed by the spin-orbit coupling.
The decay contributions from the excited states are taken into account in the final analysis in order to compare to the experimental data on $D^0$, $D_s$ and $\Lambda_c$.

\section{Numerical results}

In this section, we present the numerical results for charmed hadron chemistry (in terms of the $D_s/D^0$ and $\Lambda_c/D^0$ ratios) and the elliptic flow $v_2$ of $D^0$, $D_s$ and $\Lambda_c$ in Pb+Pb collisions at 5.02~TeV, Xe+Xe collisions at 5.44~TeV, Ar+Ar collisions at 5.85~TeV and O+O collisions at 6.5~TeV.

\begin{figure}[tbh]
\includegraphics[width=0.89\linewidth]{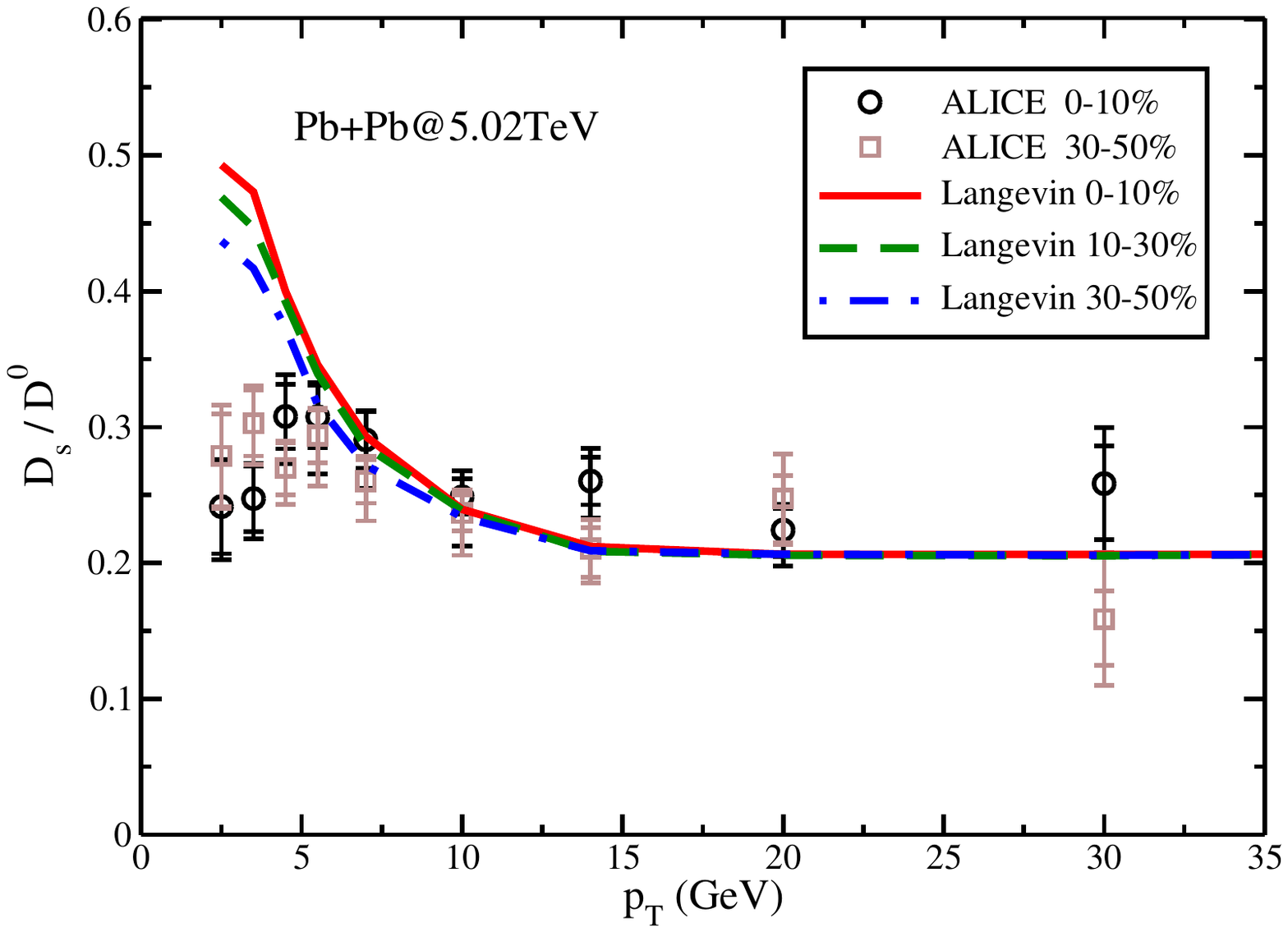}
\includegraphics[width=0.89\linewidth]{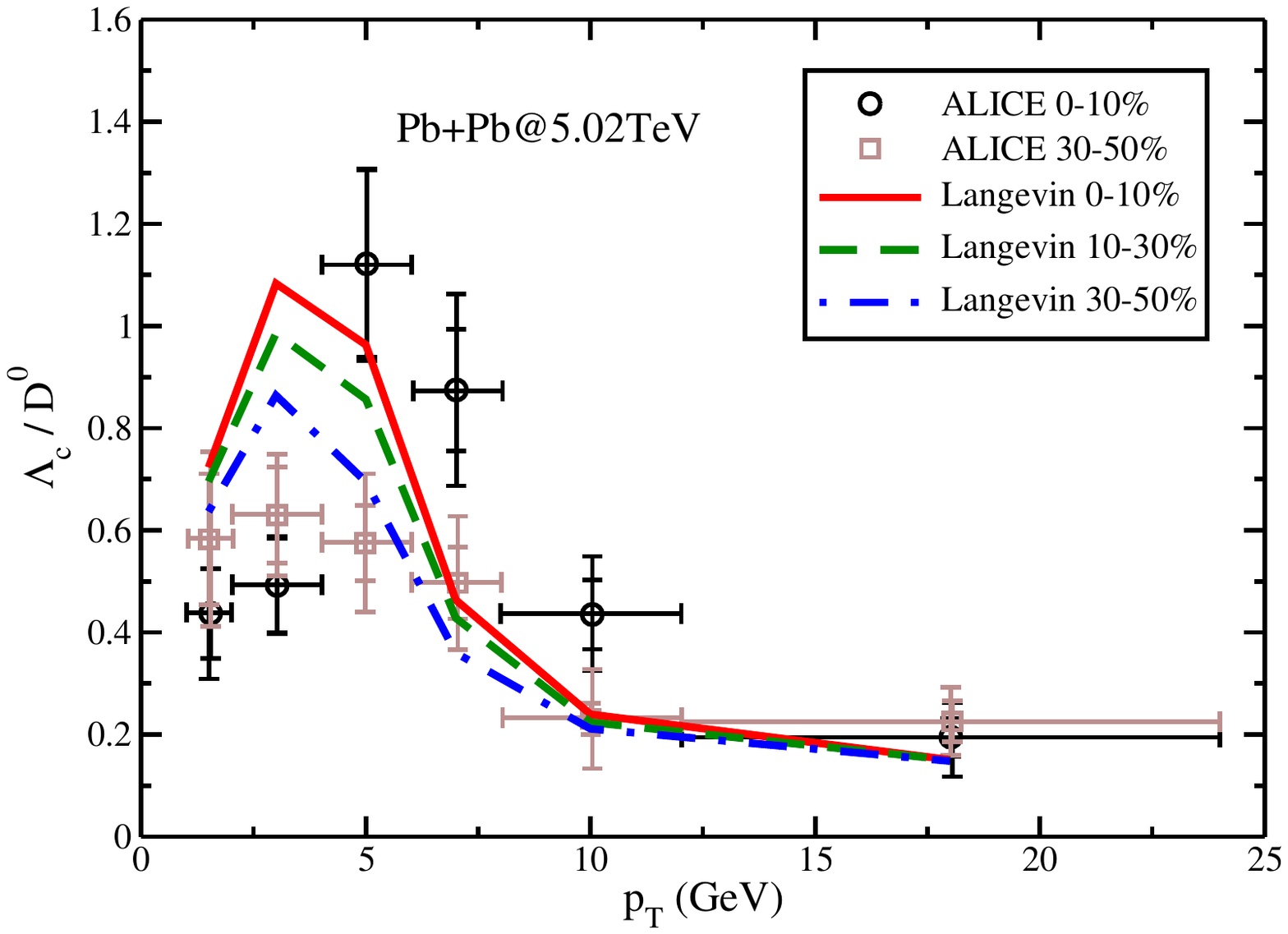}
\caption{The $D_s/D^0$ and $\Lambda_c/D^0$ ratios as a function of $p_T$ in central 0-10\%, 10-30\% and 30-50\% Pb+Pb collisions at $\sqrt{s_{NN}}=5.02$~TeV. The ALICE data $D_s/D^0$ and $\Lambda_c/D^0$ ratios in central 0-10\% and 30-50\% Pb+Pb collisions~\cite{Vermunt:2019ecg,ALICE:2021bib,ALICE:2021kfc} are shown for comparison.}
\label{fig_ratio_PbPb}
\end{figure}

\begin{figure}[tbh]
\includegraphics[width=0.89\linewidth]{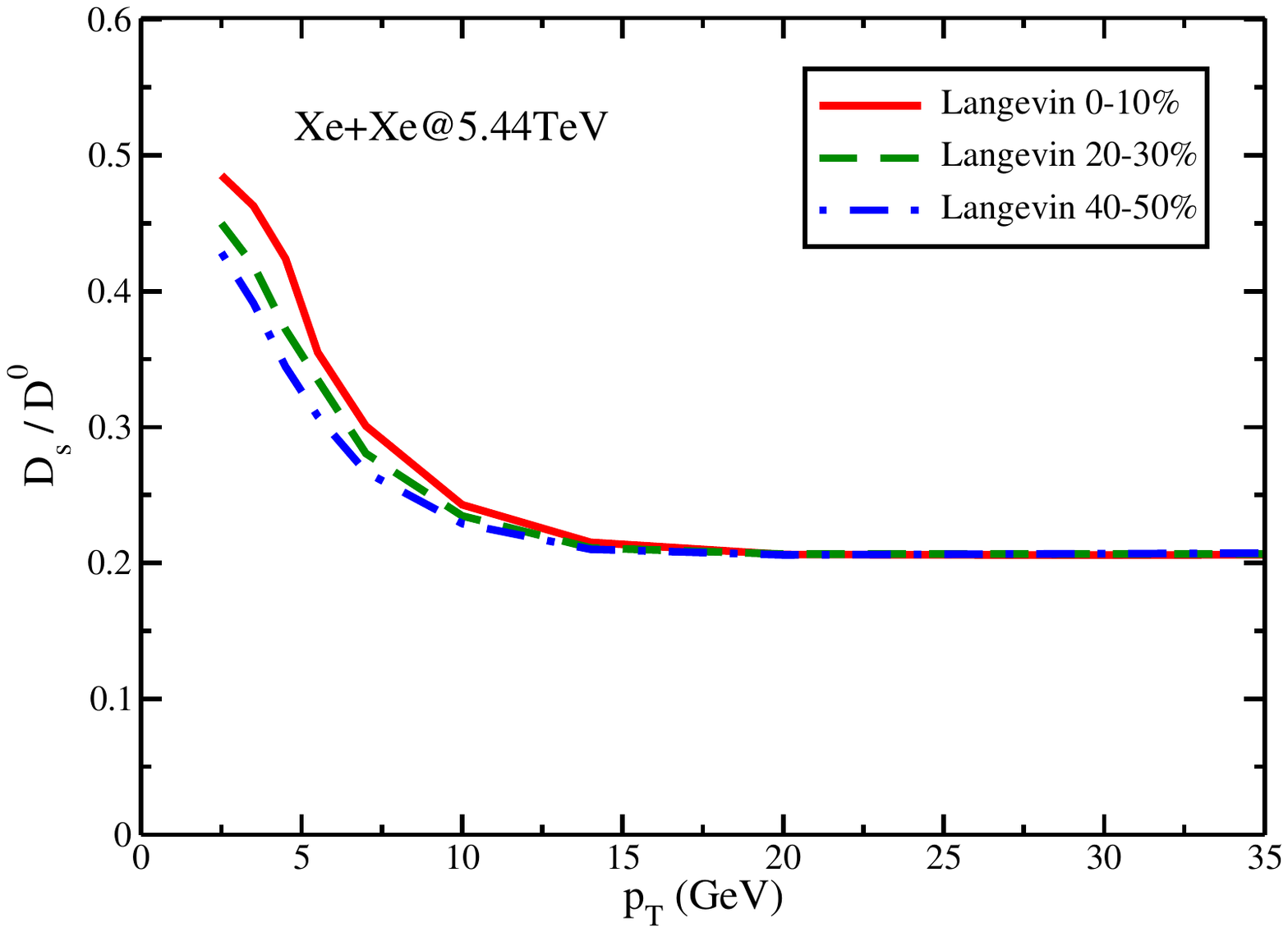}
\includegraphics[width=0.89\linewidth]{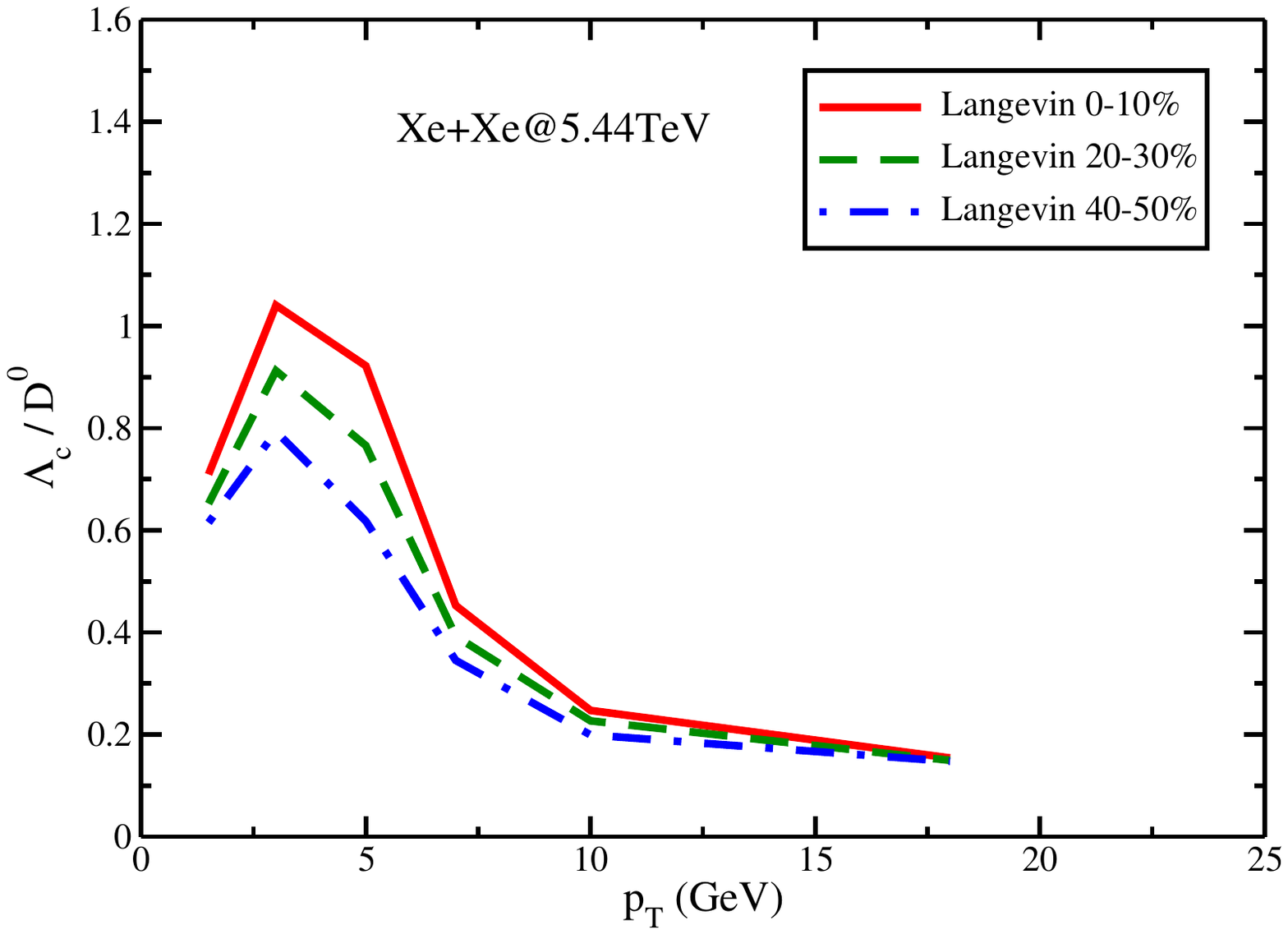}
\caption{The $D_s/D^0$ and $\Lambda_c/D^0$ ratios as a function of $p_T$ in central 0-10\%, 20-30\% and 40-50\% Xe+Xe collisions at $\sqrt{s_{NN}}=5.44$~TeV.} \label{fig_ratio_XeXe}
\label{fig_ratio_XeXe}
\end{figure}

\begin{figure}[tbh]
\includegraphics[width=0.89\linewidth]{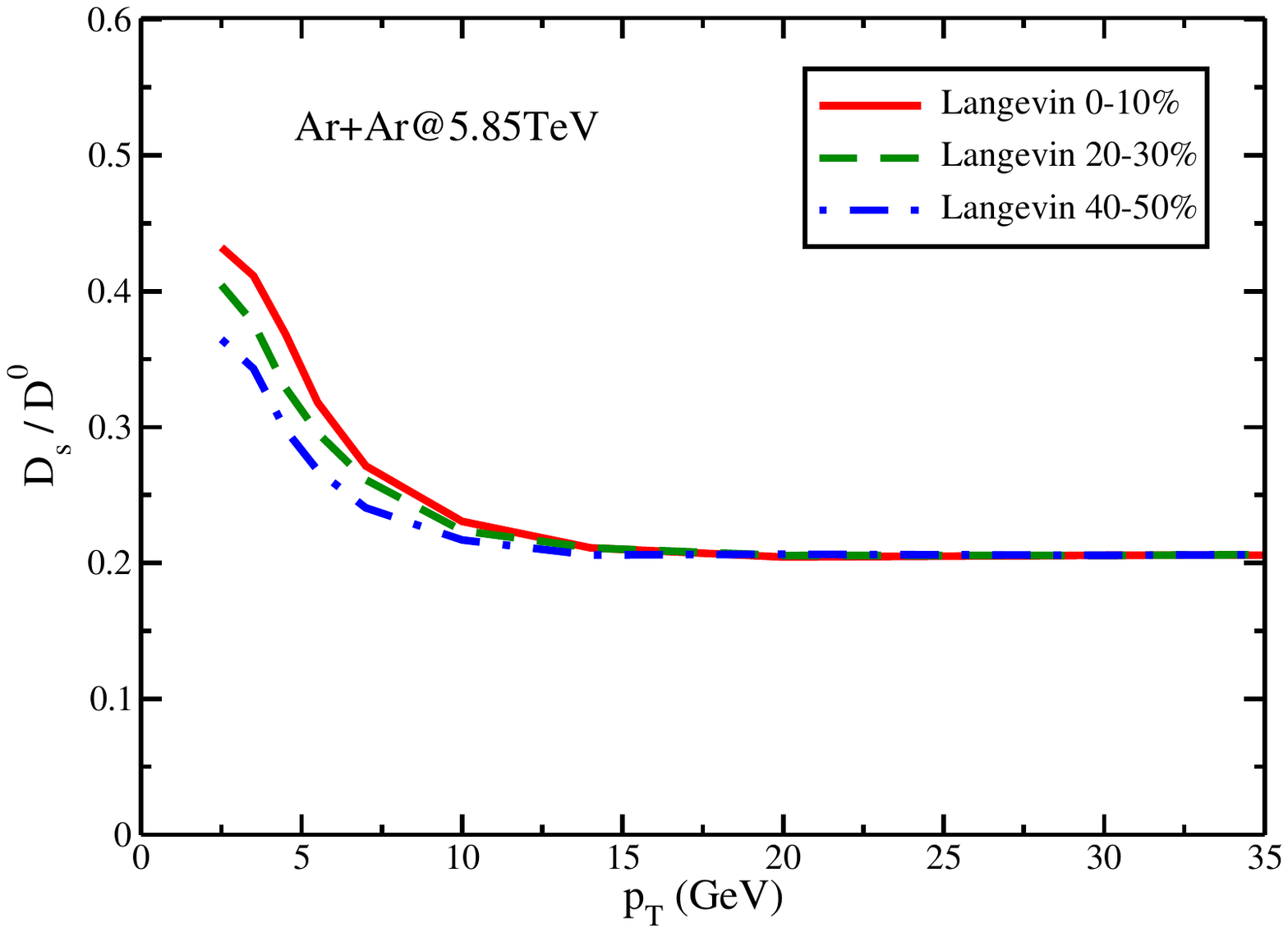}
\includegraphics[width=0.89\linewidth]{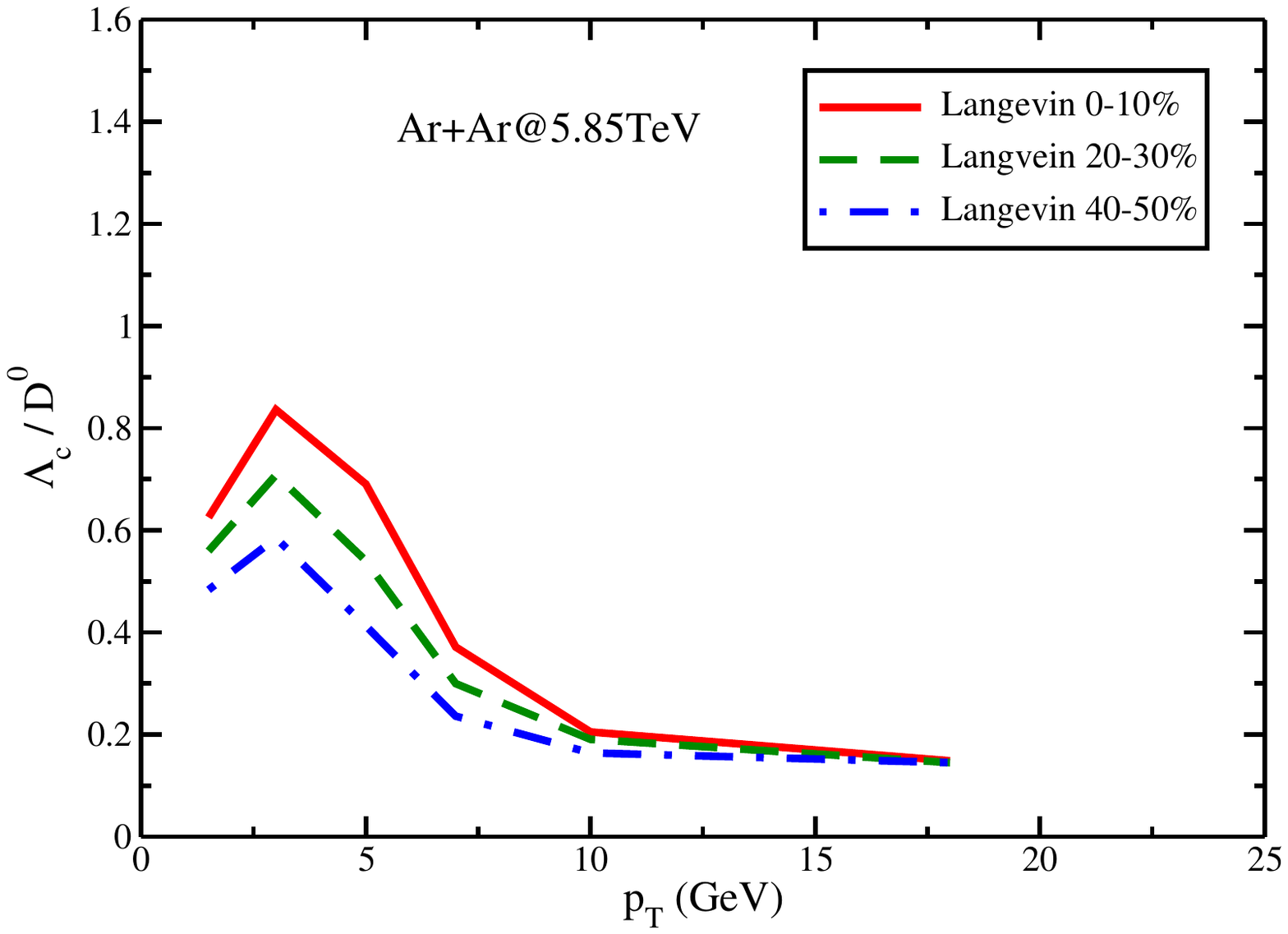}
\caption{The $D_s/D^0$ and $\Lambda_c/D^0$ ratios as a function of $p_T$ in central 0-10\%, 20-30\% and 40-50\% Ar+Ar collisions at $\sqrt{s_{NN}}=5.85$~TeV.} \label{fig_ratio_ArAr}
\label{fig_ratio_ArAr}
\end{figure}

\begin{figure}[tbh]
\includegraphics[width=0.89\linewidth]{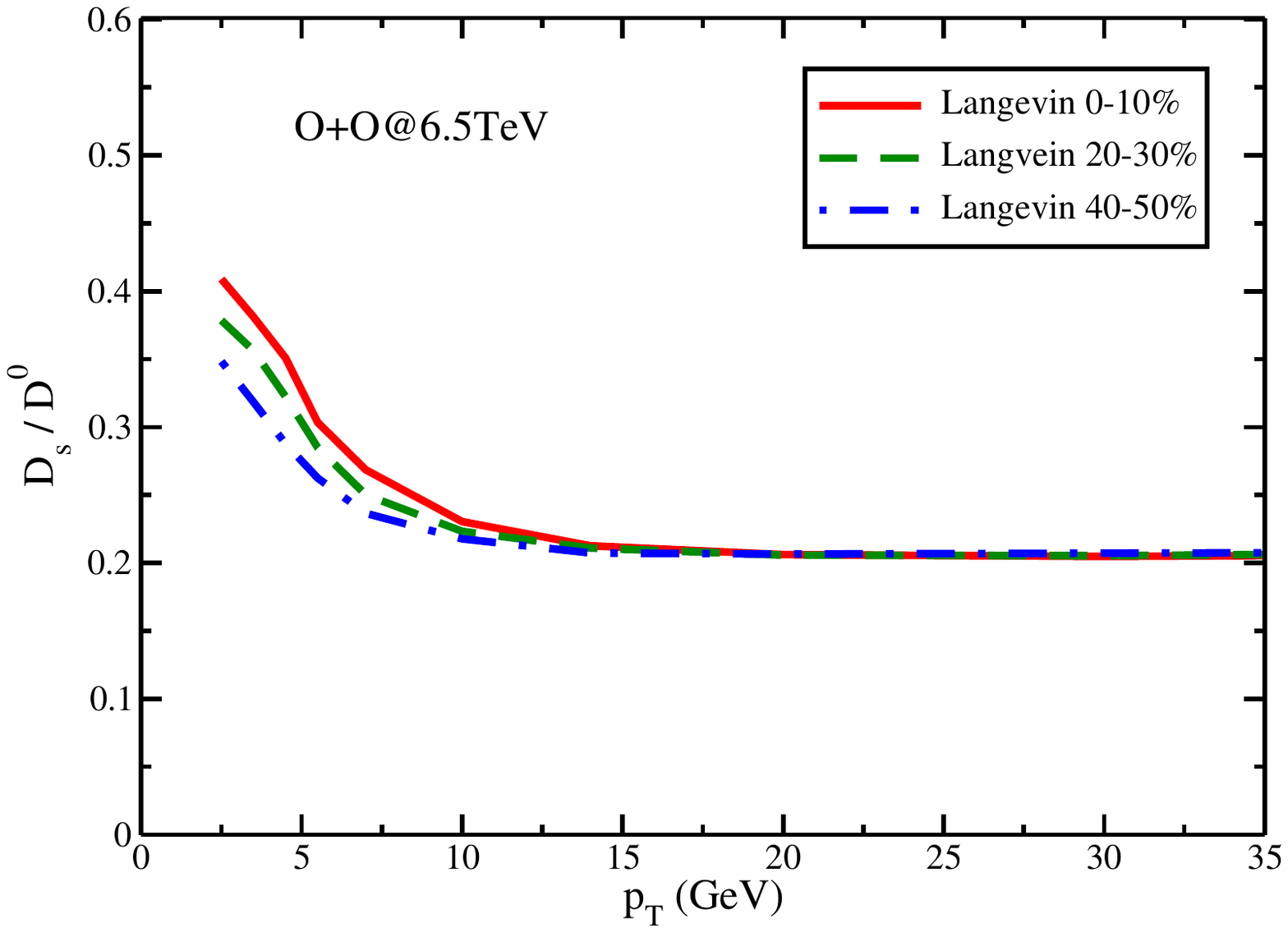}
\includegraphics[width=0.89\linewidth]{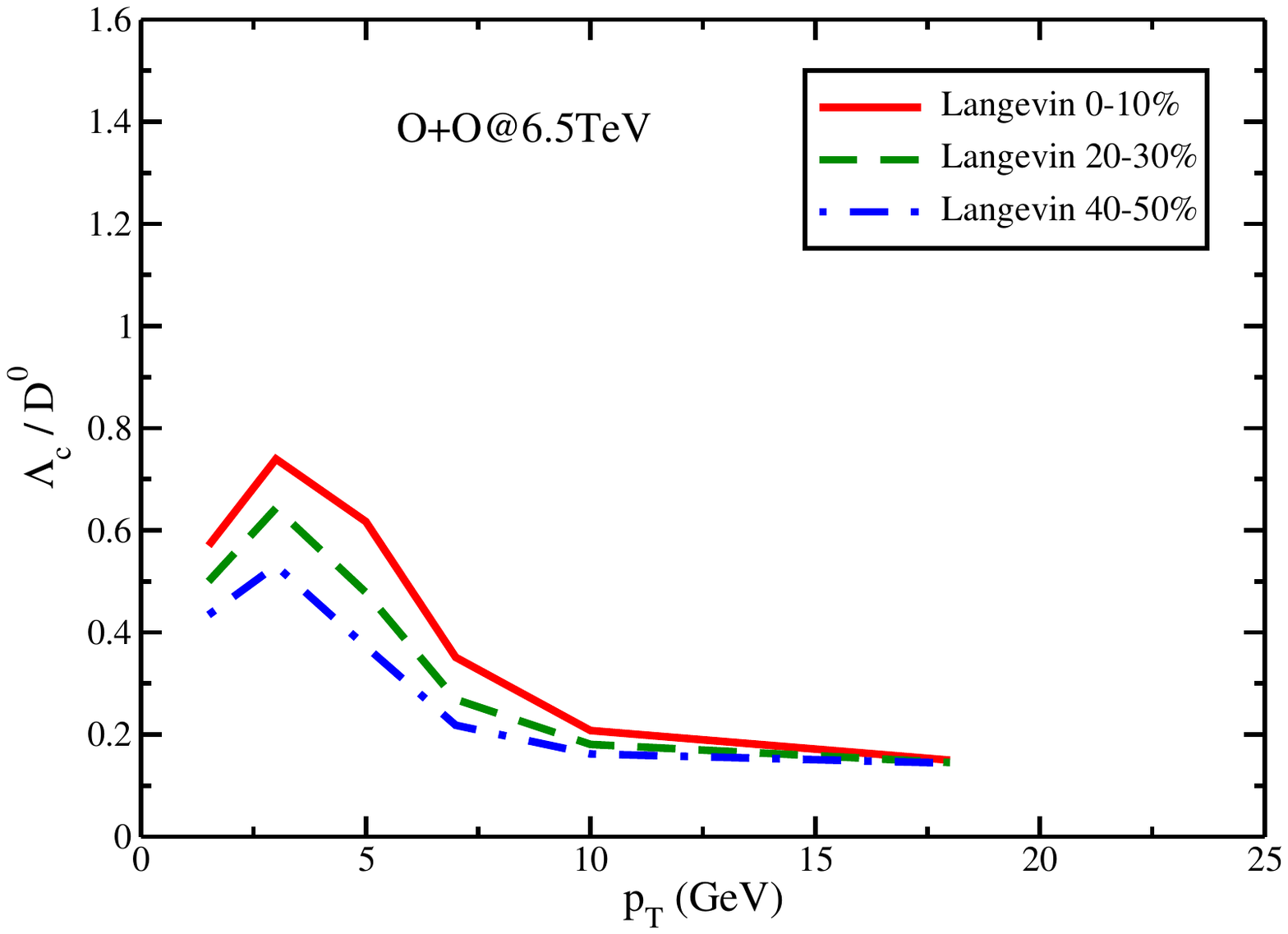}
\caption{The $D_s/D^0$ and $\Lambda_c/D^0$ ratios as a function of $p_T$ in central 0-10\%, 20-30\% and 40-50\% Ar+Ar collisions at $\sqrt{s_{NN}}=6.5$~TeV.} \label{fig_ratio_OO}
\label{fig_ratio_OO}
\end{figure}

\begin{figure}[tbh]
\includegraphics[width=0.89\linewidth]{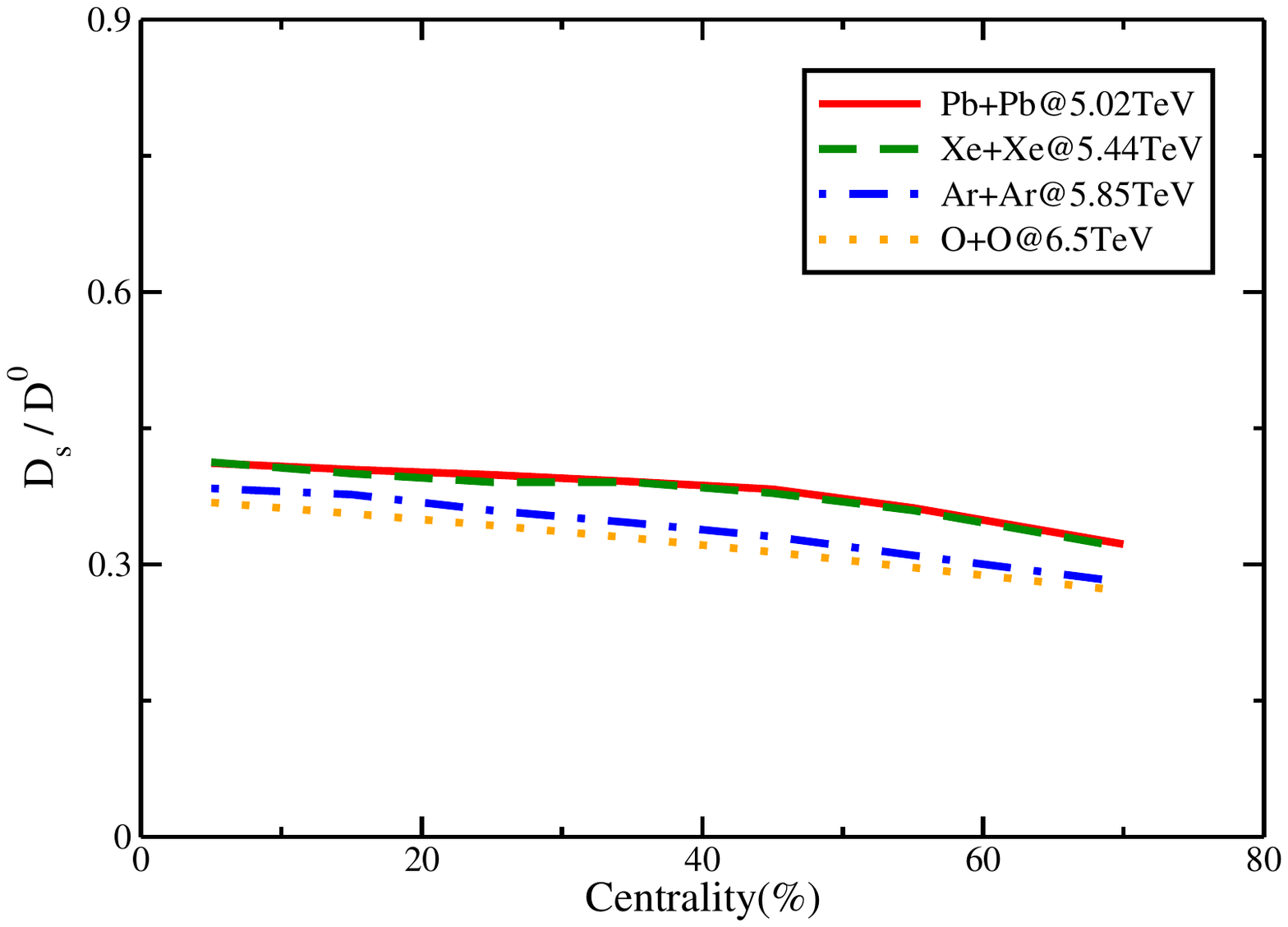}
\includegraphics[width=0.89\linewidth]{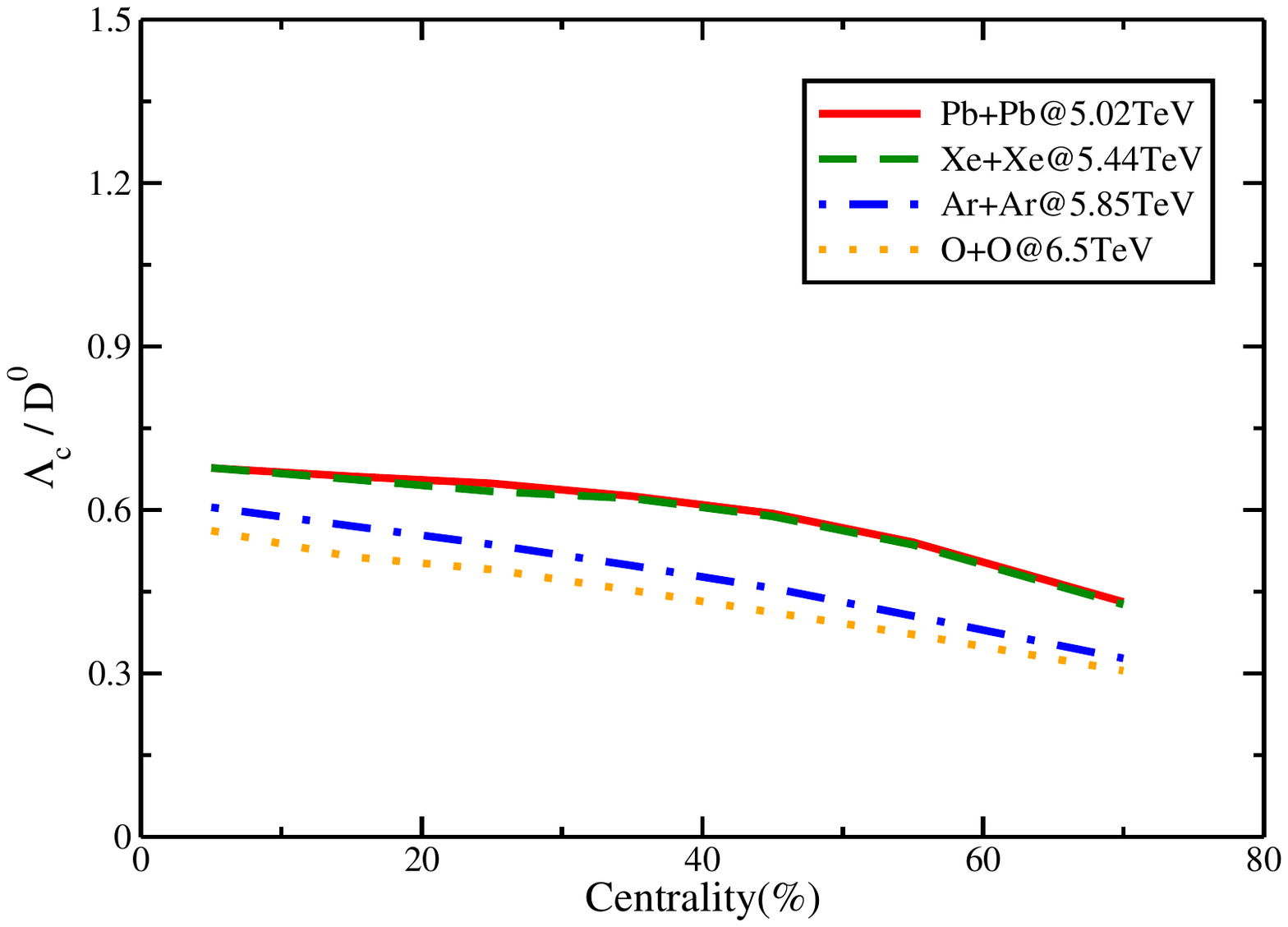}
\caption{The $D_s/D^0$ and $\Lambda_c/D^0$ ratios as a function of centrality in Pb+Pb, Xe+Xe, Ar+Ar and O+O collisions at the LHC energies.}
\label{fig_ratio_centrality}
\end{figure}

\subsection{Charmed hadron chemistry}

We first shown in Fig.~\ref{fig_ratio_PbPb} the charmed hadron ratios $D_s/D^0$ and $\Lambda_c/D^0$ as a function of $p_T$ in Pb+Pb collisions at $\sqrt{s_{NN}}=5.02$~TeV. The results are shown for three different centralities (0-10\%, 10-30\% and 30-50\%) from our model calculations and two centralities (0-10\% and 30-50\%) from the ALICE data~\cite{Vermunt:2019ecg,ALICE:2021bib,ALICE:2021kfc}.
One can see that both $D_s/D^0$ and $\Lambda_c/D^0$ ratios increase from peripheral to central collisions, as shown by both our model and the ALICE data.
Such centrality dependence is clearly the signature of the hot medium effect. Because of the strangeness enhancement~\cite{Rafelski:1982pu,Koch:1986ud} and parton coalescence effects, $D_s/D^0$ and $\Lambda_c/D^0$ ratios are higher in more central collisions where the QGP medium is larger and hotter.
We also observe that the centrality dependence is stronger for $\Lambda_c/D^0$ than $D_s/D^0$. This is due to the baryon-to-meson enhancement effect again resulting from the coalescence of thermal partons from the hot QGP medium~\cite{Fries:2003vb, Greco:2003xt}.

Figure~\ref{fig_ratio_XeXe} shows our prediction for the $D_s/D^0$ and $\Lambda_c/D^0$ ratios as a function of $p_T$ in Xe+Xe collisions $\sqrt{s_{NN}}=5.44$~TeV for three centralities (0-10\%, 20-30\% and 40-50\%). Compared to our model calculations in Pb+Pb collisions, one can see that the magnitudes of two charmed hadron ratios  $D_s/D^0$ and $\Lambda_c/D^0$  in Xe+Xe collisions are similar to those in Pb+Pb collisions.
This means that the system size does not play very important roles between Pb+Pb and Xe+Xe collisions.
One may understand this as follows. While Xe+Xe systems are indeed smaller than Pb+Pb systems, the QGP produced in Xe+Xe collisions are still large enough to develop the radial flow, which plays important roles in heavy flavor hadron production via parton coalescence mechanism~\cite{Cao:2019iqs}.

Figures~\ref{fig_ratio_ArAr} and \ref{fig_ratio_OO} show our predictions for the $D_s/D^0$ and $\Lambda_c/D^0$ ratios as a function of $p_T$ in Ar+Ar collisions at $\sqrt{s_{NN}}=5.85$~TeV and in O+O collisions at $\sqrt{s_{NN}}=6.5$~TeV for three centralities (0-10\%, 10-20\% and 40-50\%).
Similar to Pb+Pb and Xe+Xe collisions, we also observe strong centrality dependence for both $D_s/D^0$ and $\Lambda_c/D^0$ ratios in Ar+Ar and O+O collisions, and the centrality dependence is stronger for $\Lambda_c/D^0$ than $D_s/D^0$.
Such strong centrality dependence can be regarded as the ``evidence" for the QGP formation in light ion collisions once confirmed by the future experiments.
Meanwhile, due to much smaller system sizes in Ar+Ar and O+O collisions, we observe that the values of $D_s/D^0$ and $\Lambda_c/D^0$ ratios in these two light ion collisions are much smaller compared to Pb+Pb collisions, though the collision energies are somewhat higher in Ar+Ar and O+O collisions.
This situation is quite different from comparing Xe+Xe collisions and Pb+Pb collisions, where the system sizes are large enough and do not play very important role.

In order to have a more direct comparison among different collision systems, we show in Fig.~\ref{fig_ratio_centrality} the charmed hadron ratios $D_s/D^0$ and $\Lambda_c/D^0$ (integrated over $p_T$) as a function of collision centrality in Pb+Pb, Xe+Xe, Ar+Ar and O+O collisions.
One can see that the  $D_s/D^0$ and $\Lambda_c/D^0$ ratios are similar in two heavy-ion collisions (Pb+Pb and Xe+Xe), while for light ion collisions (Ar+Ar and O+O), the $D_s/D^0$ and $\Lambda_c/D^0$ ratios are much smaller compared to Pb+Pb and Xe+Xe collisions due to the significant system size effect.
Another important observation is the strong centrality dependence for the $D_s/D^0$ and $\Lambda_c/D^0$ ratios observed in all four collision systems.
In addition, the centrality dependence is stronger for $\Lambda_c/D^0$ than $D_s/D^0$ due to the baryon-to-meson enhancement effect.
This mainly originates from the effect of parton coalescence, which have been regarded as the evidences for the formation of QGP in Au+Au collisions at RHIC and Pb+Pb collisions at the LHC.
Therefore, once the above phenomenon is observed in future O+O collisions, it will serve as a strong evidence for the formation of QGP in O+O collisions.

\begin{figure}[tbh]
\includegraphics[width=0.89\linewidth]{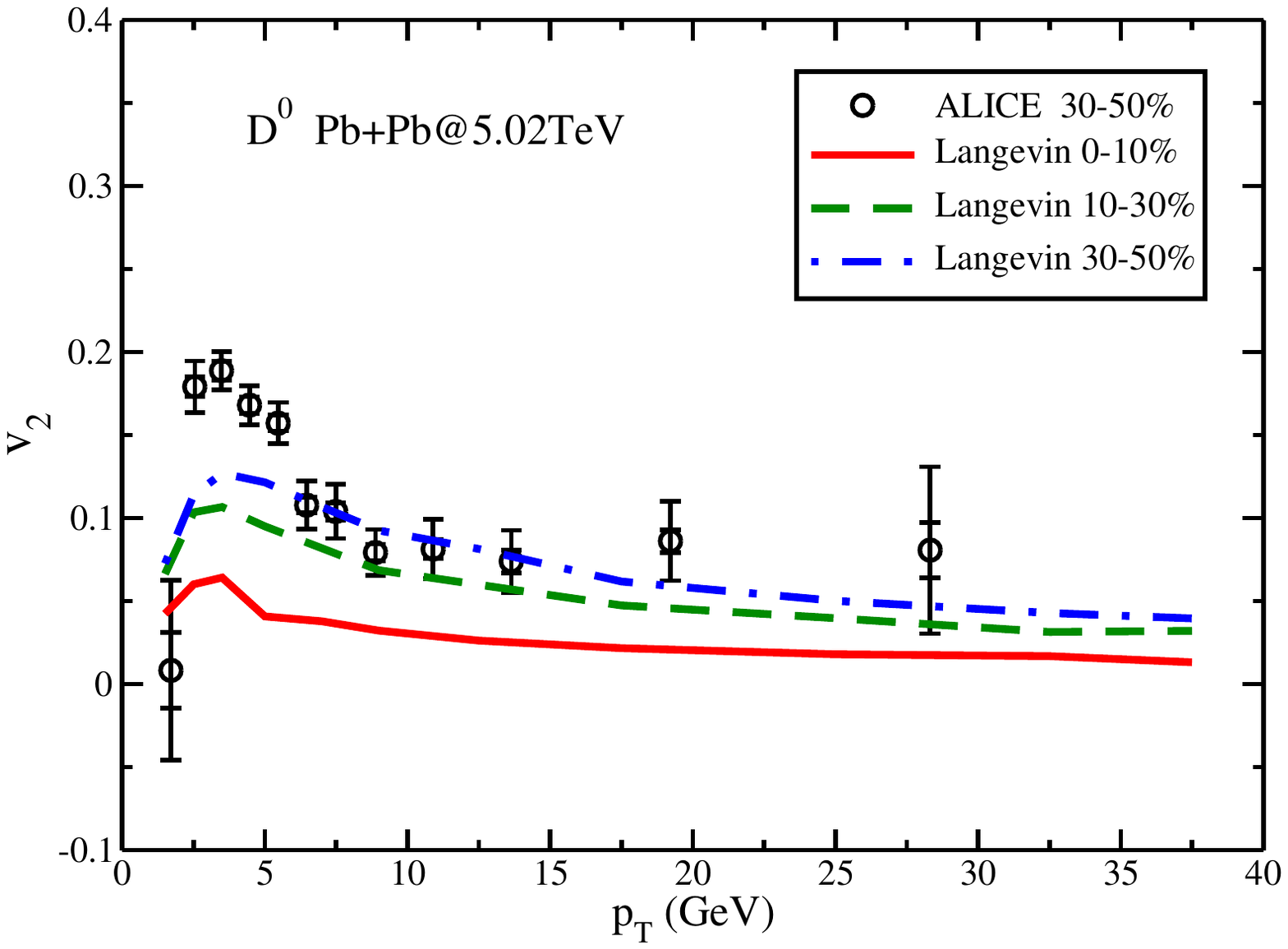}
\includegraphics[width=0.89\linewidth]{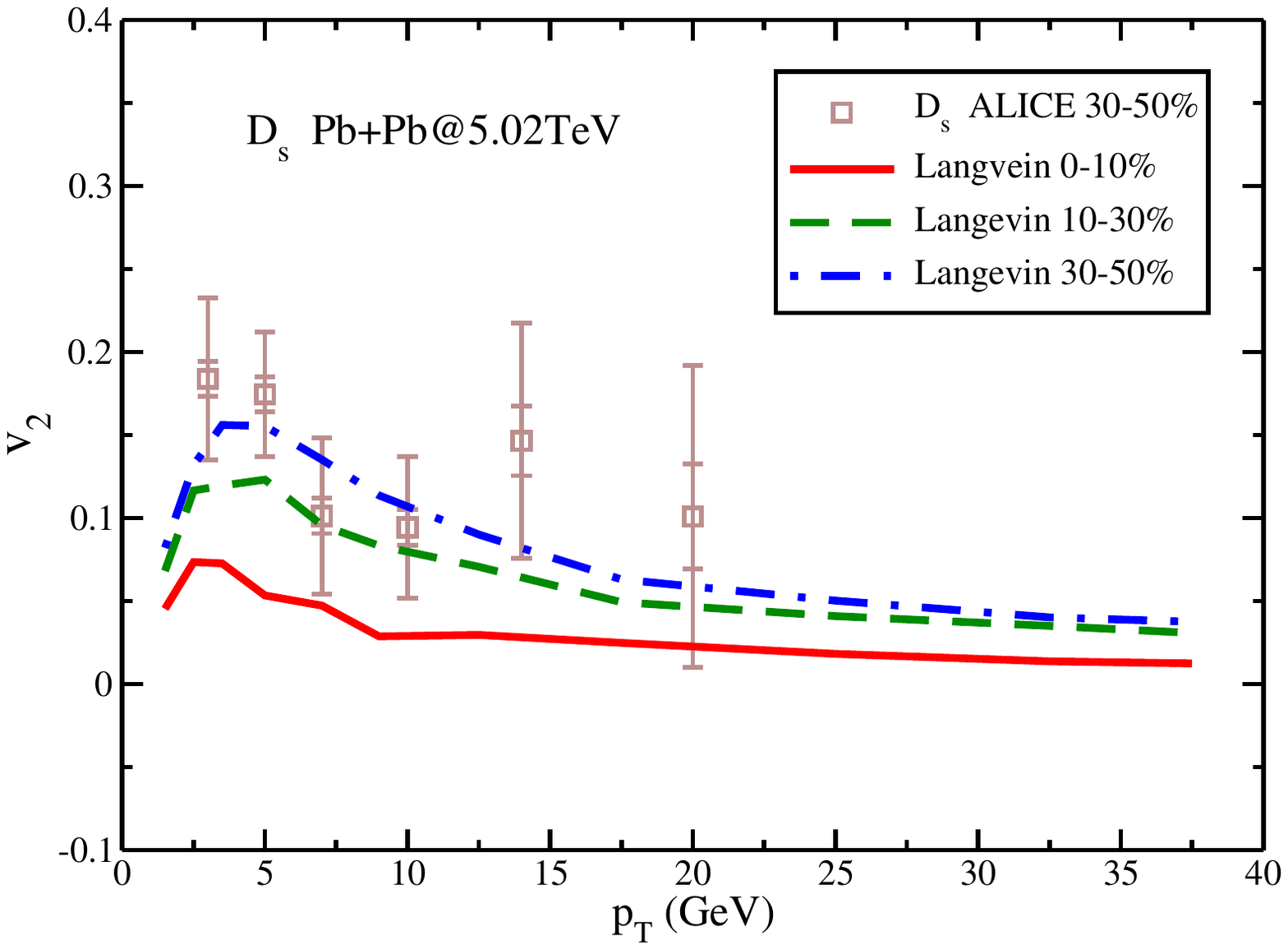}
\includegraphics[width=0.89\linewidth]{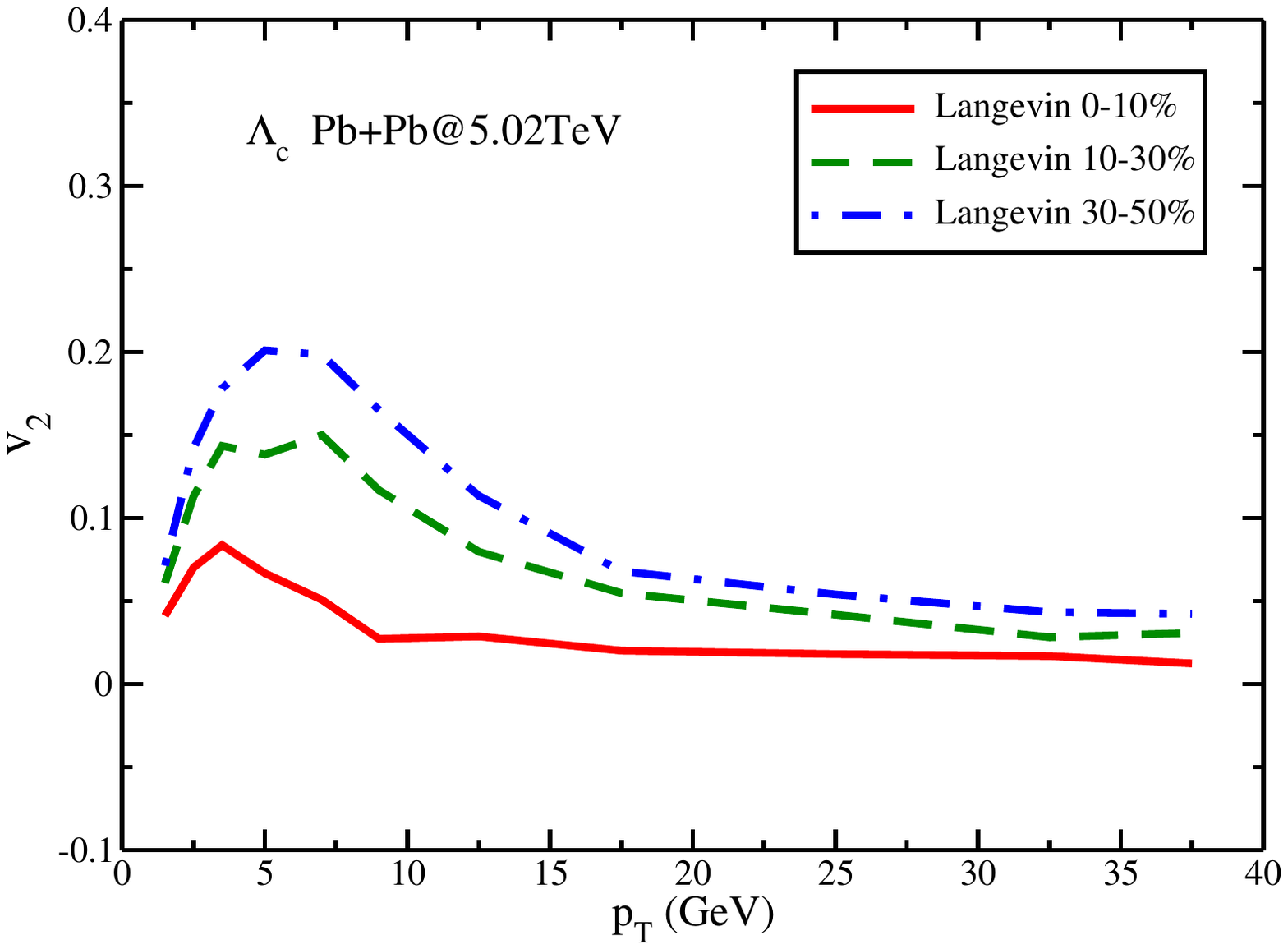}
\caption{The elliptic flow $v_2$ for $D^0$, $D_s$ and $\Lambda_c$ as a function of $p_T$ in central 0-10\%, 10-30\% and 30-50\% Pb+Pb collisions at $\sqrt{s_{NN}}=5.02$~TeV. The ALICE data on $D^0$ and $D_s$ meson $v_2$ in 30-50\% Pb+Pb collisions~\cite{Vermunt:2019ecg,ALICE:2020iug,ALICE:2021kfc} are shown for comparison.}
\label{fig_v2_PbPb}
\end{figure}

\begin{figure}[tbh]
\includegraphics[width=0.89\linewidth]{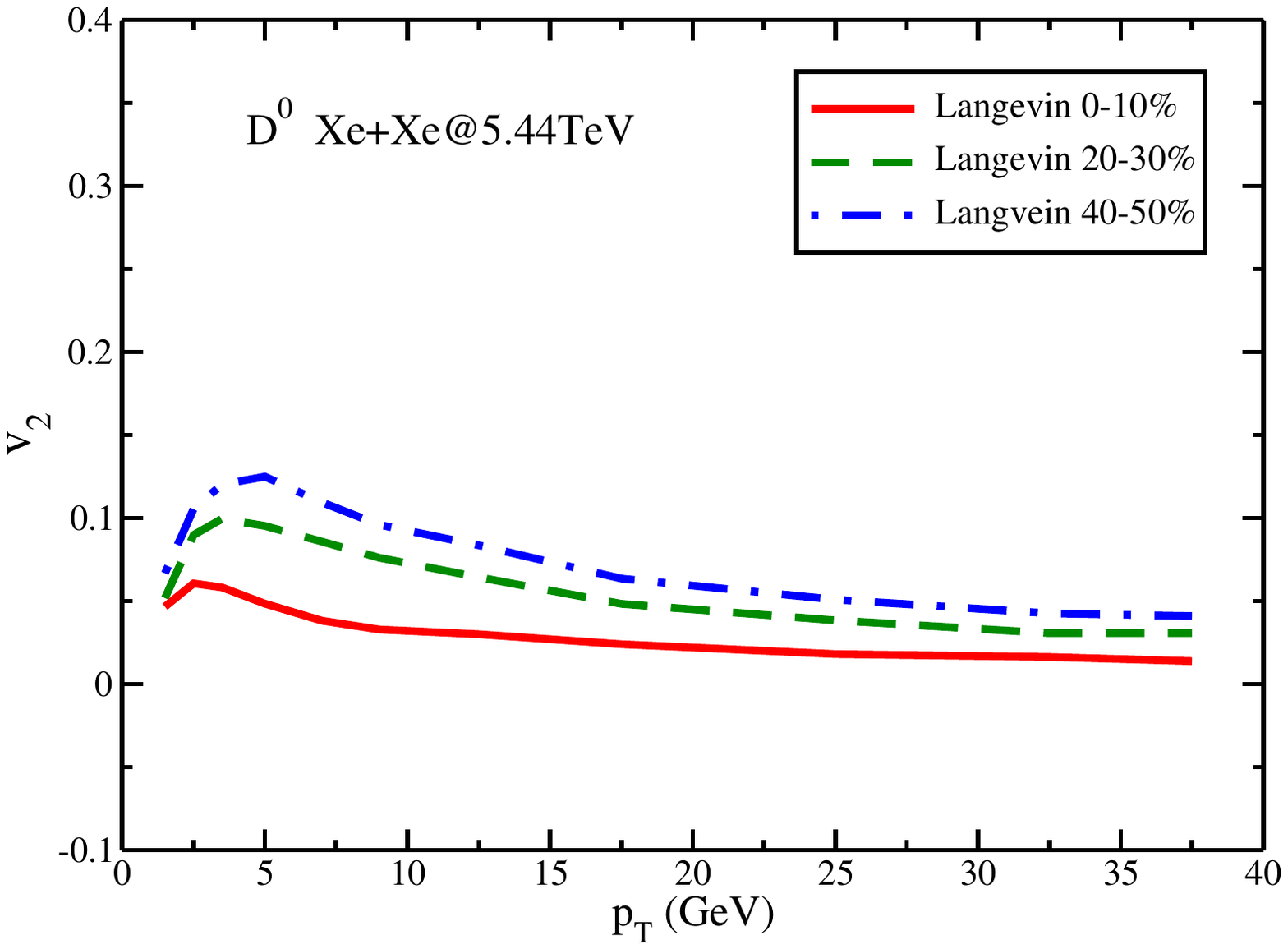}
\includegraphics[width=0.89\linewidth]{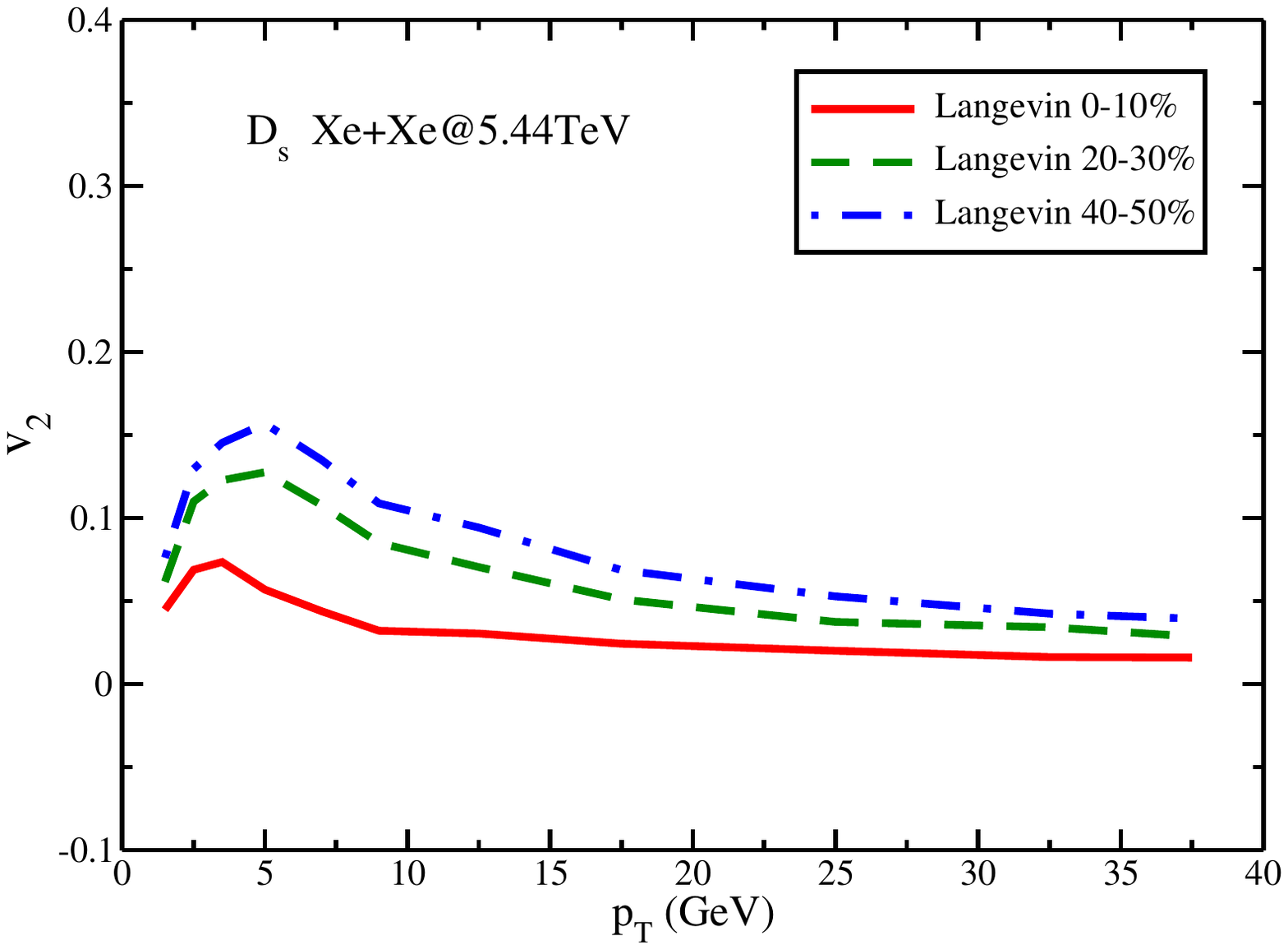}
\includegraphics[width=0.89\linewidth]{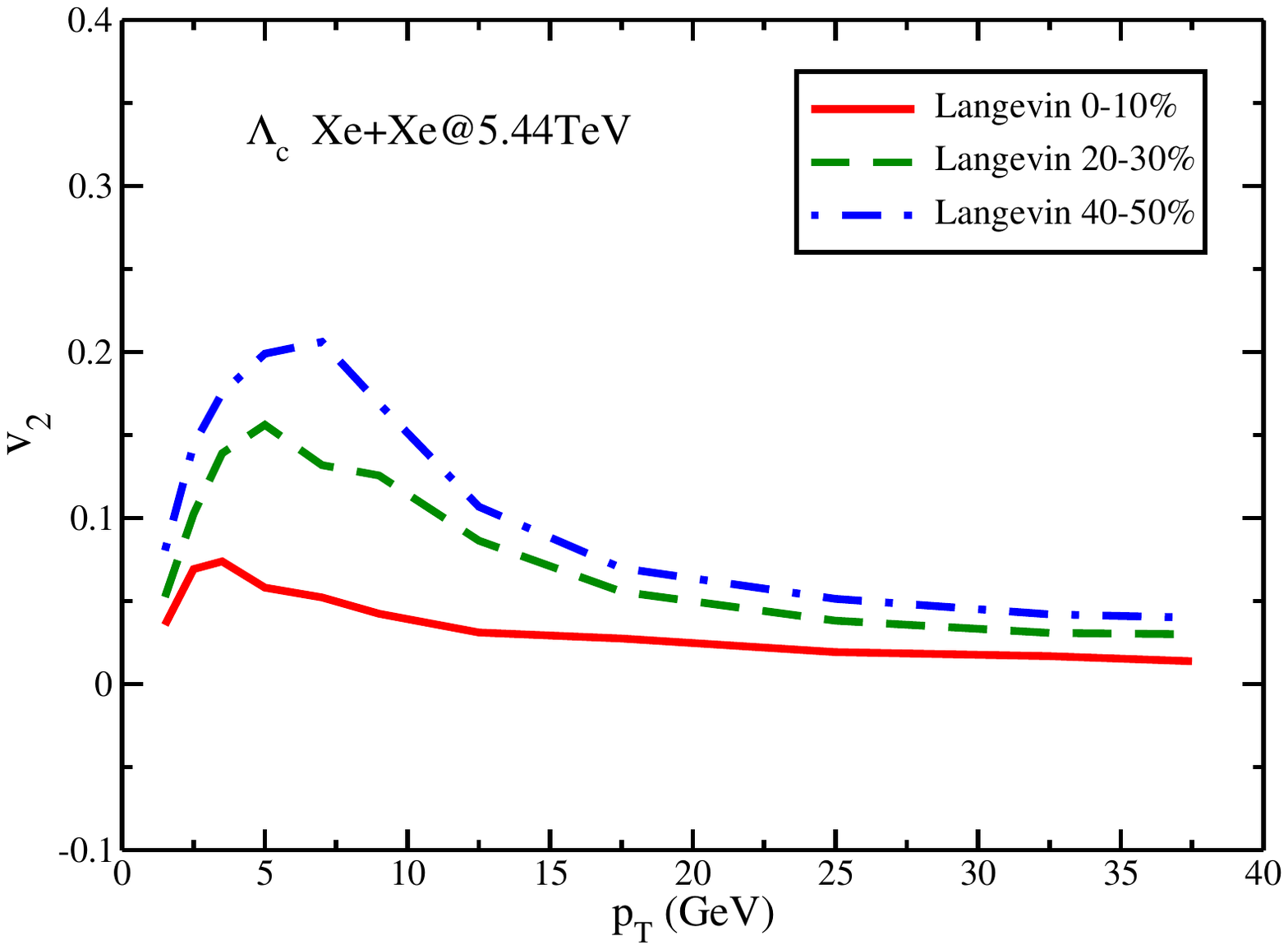}
\caption{The elliptic flow $v_2$ for $D^0$, $D_s$ and $\Lambda_c$ as a function of $p_T$ in central 0-10\%, 20-30\% and 40-50\% Xe+Xe collisions at $\sqrt{s_{NN}}=5.44$~TeV.}
\label{fig_v2_XeXe}
\end{figure}

\begin{figure}[tbh]
\includegraphics[width=0.89\linewidth]{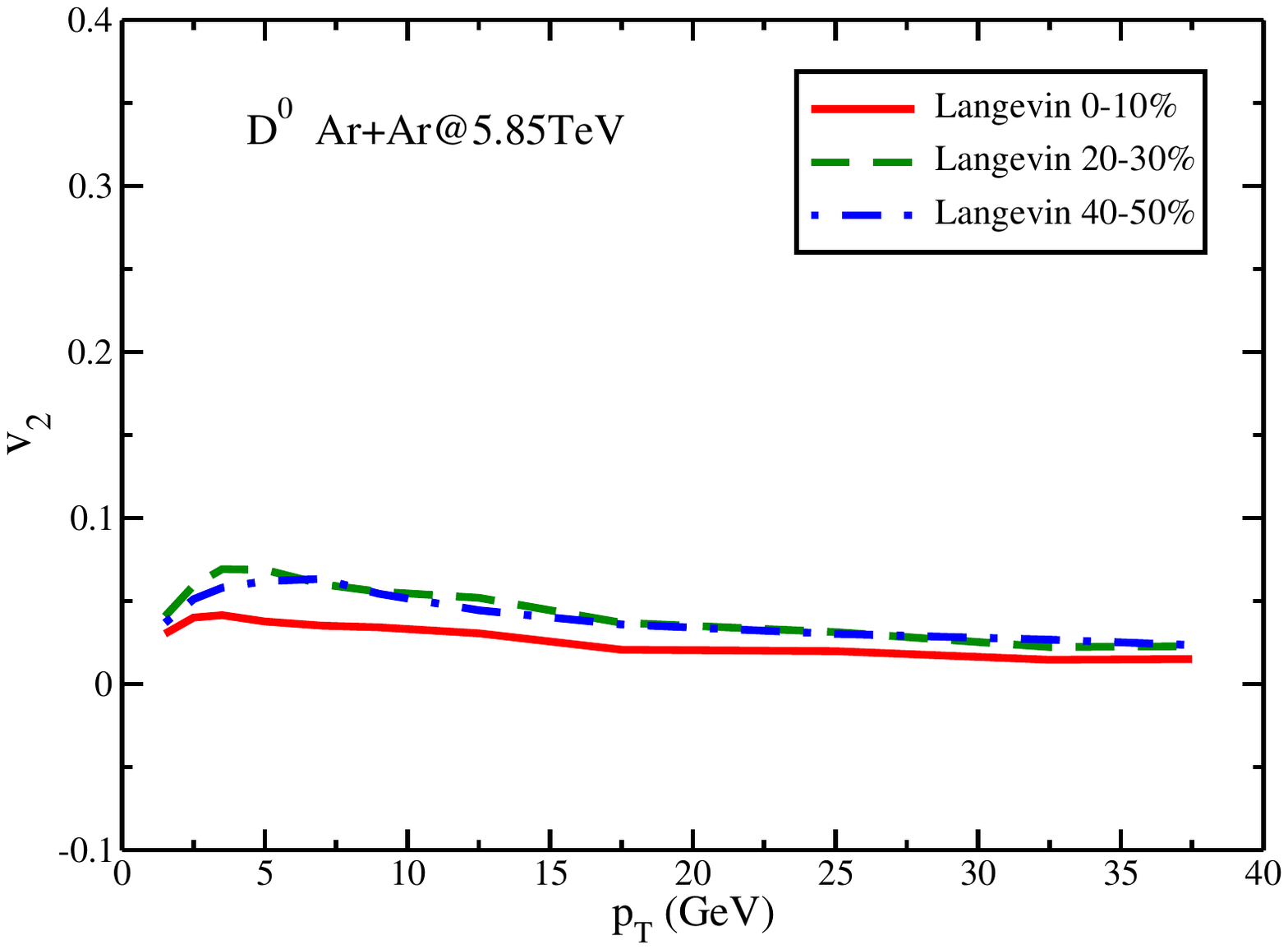}
\includegraphics[width=0.89\linewidth]{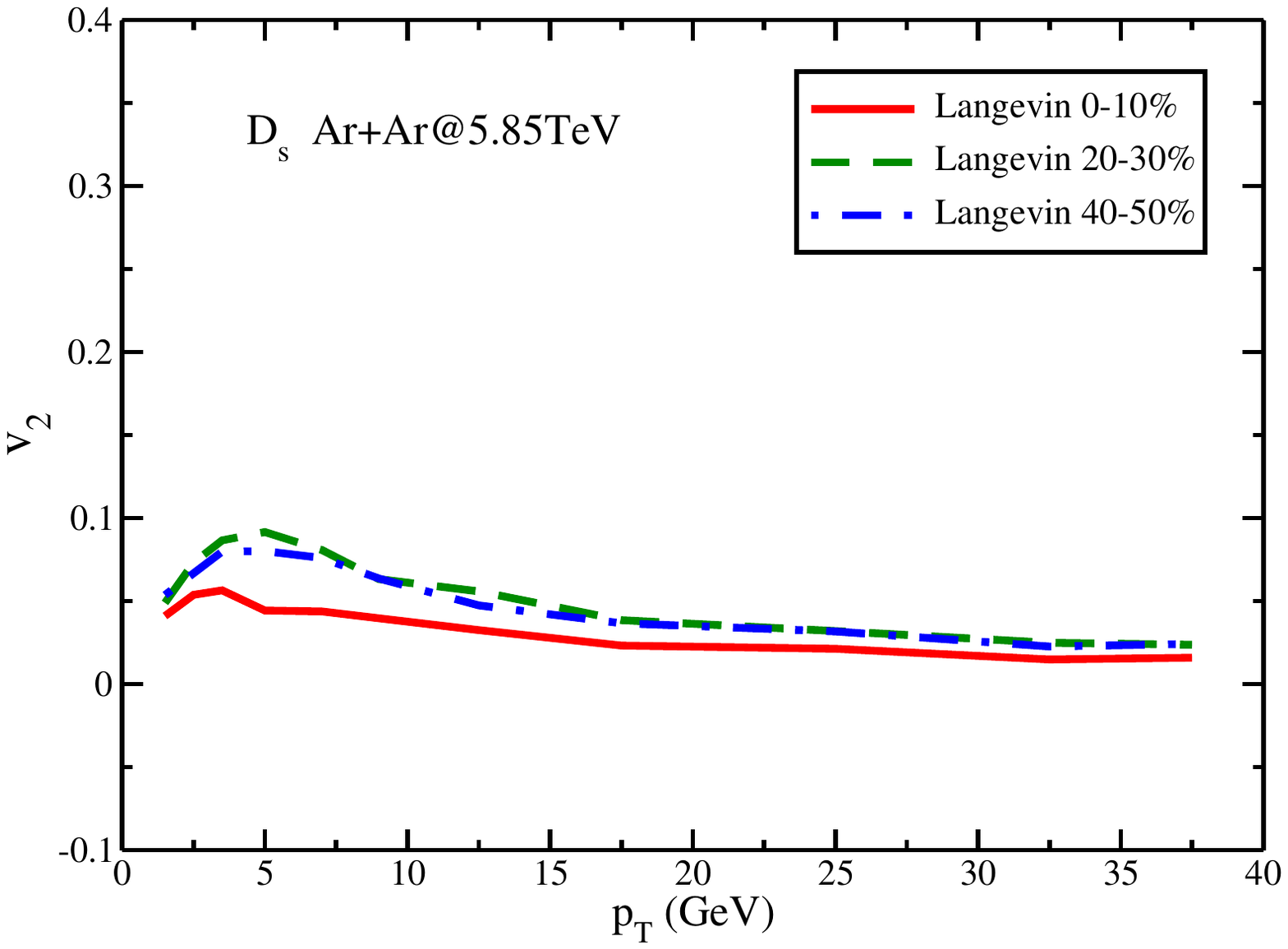}
\includegraphics[width=0.89\linewidth]{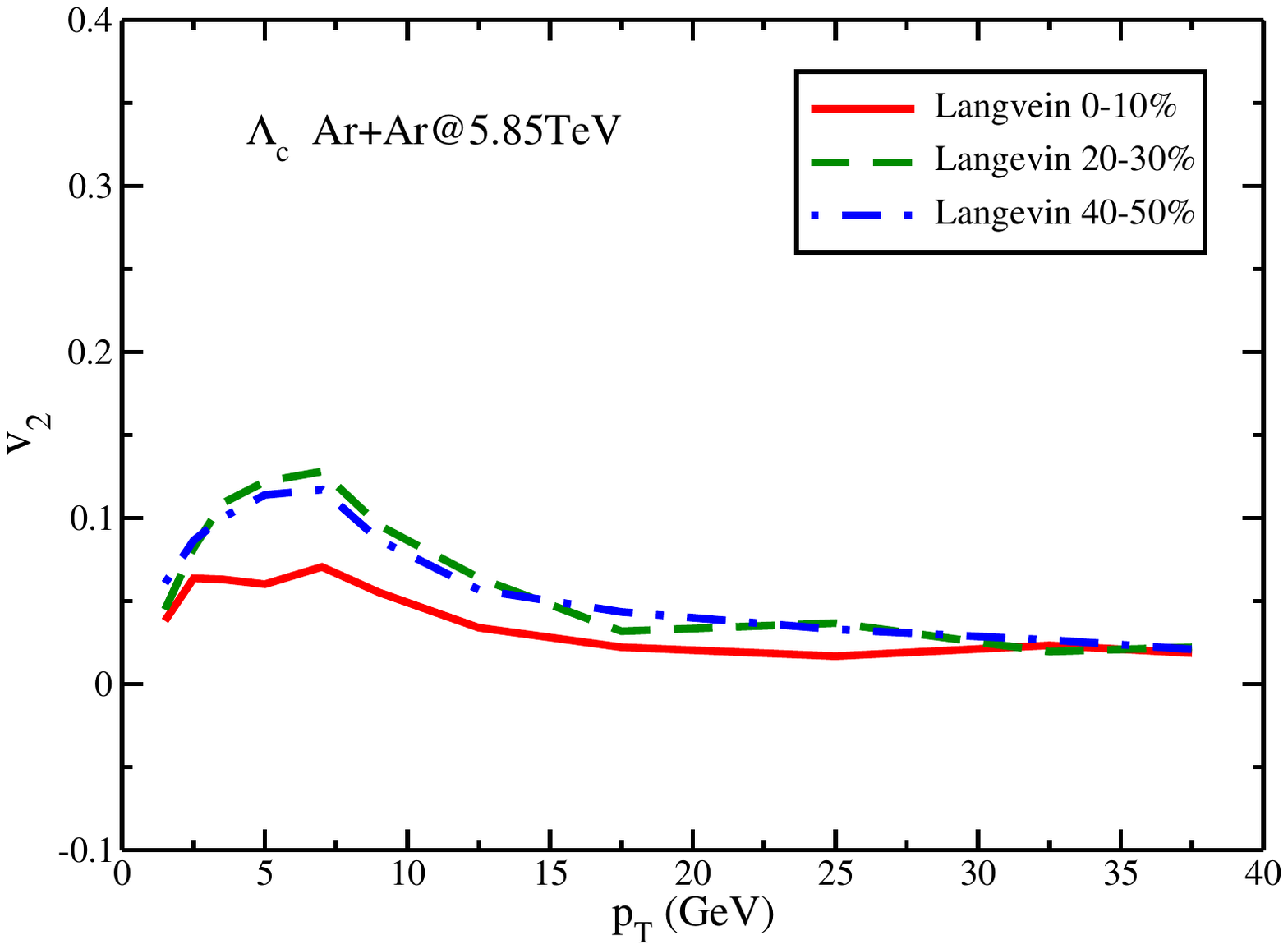}
\caption{The elliptic flow $v_2$ for $D^0$, $D_s$ and $\Lambda_c$ as a function of $p_T$ in central 0-10\%, 20-30\% and 40-50\% Ar+Ar collisions at $\sqrt{s_{NN}}=5.86$~TeV.}
\label{fig_v2_ArAr}
\end{figure}

\begin{figure}[tbh]
\includegraphics[width=0.89\linewidth]{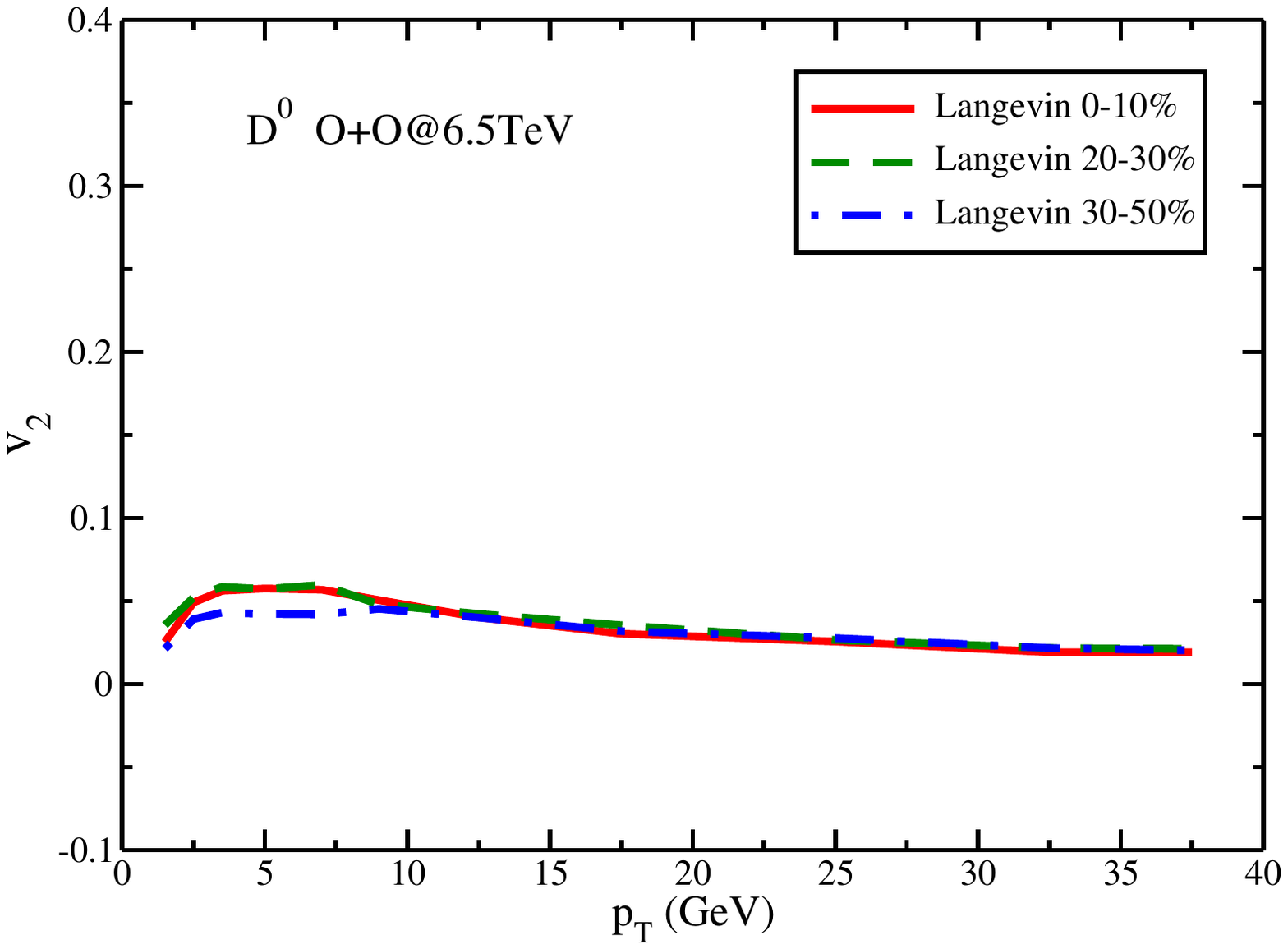}
\includegraphics[width=0.89\linewidth]{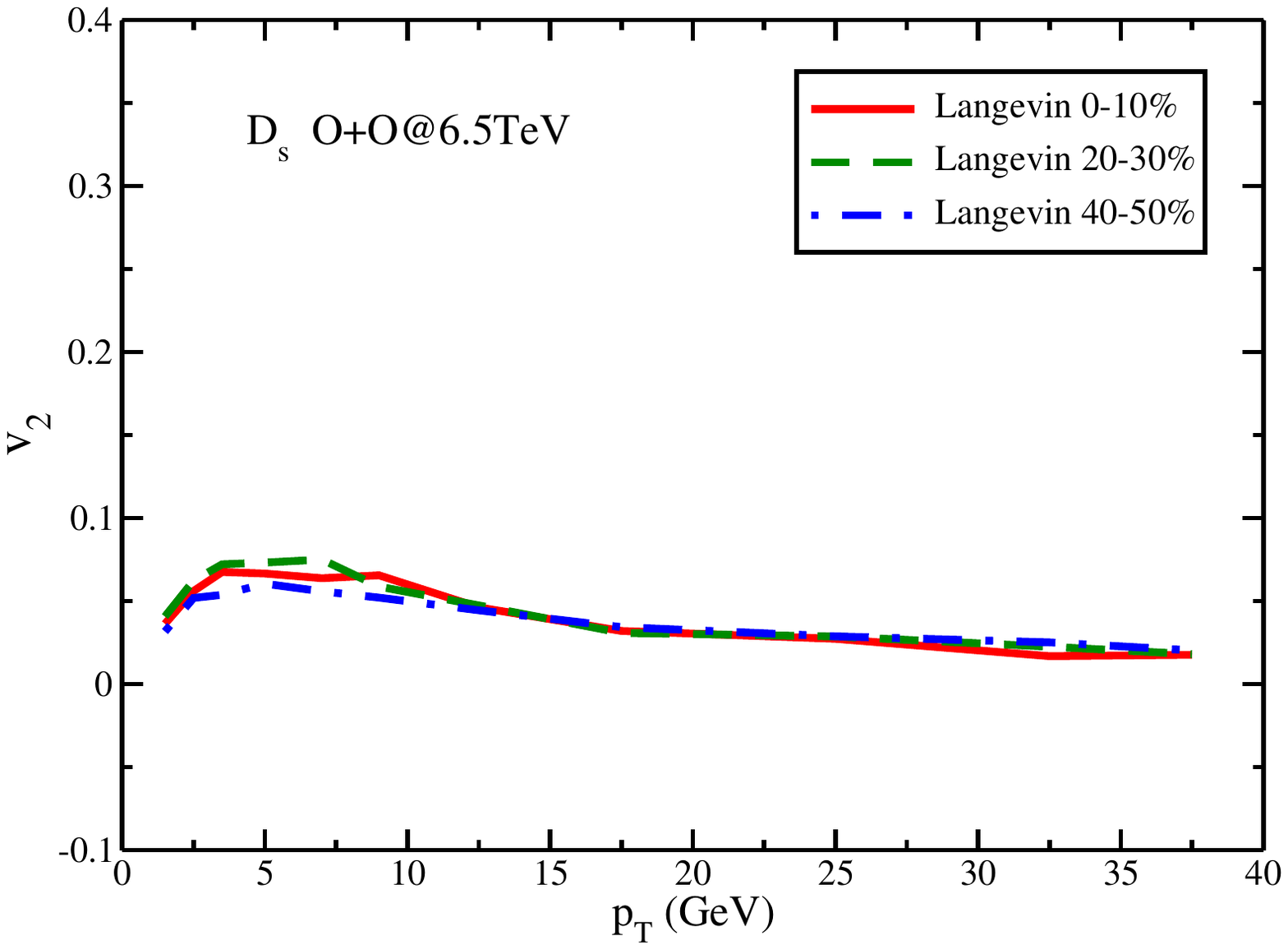}
\includegraphics[width=0.89\linewidth]{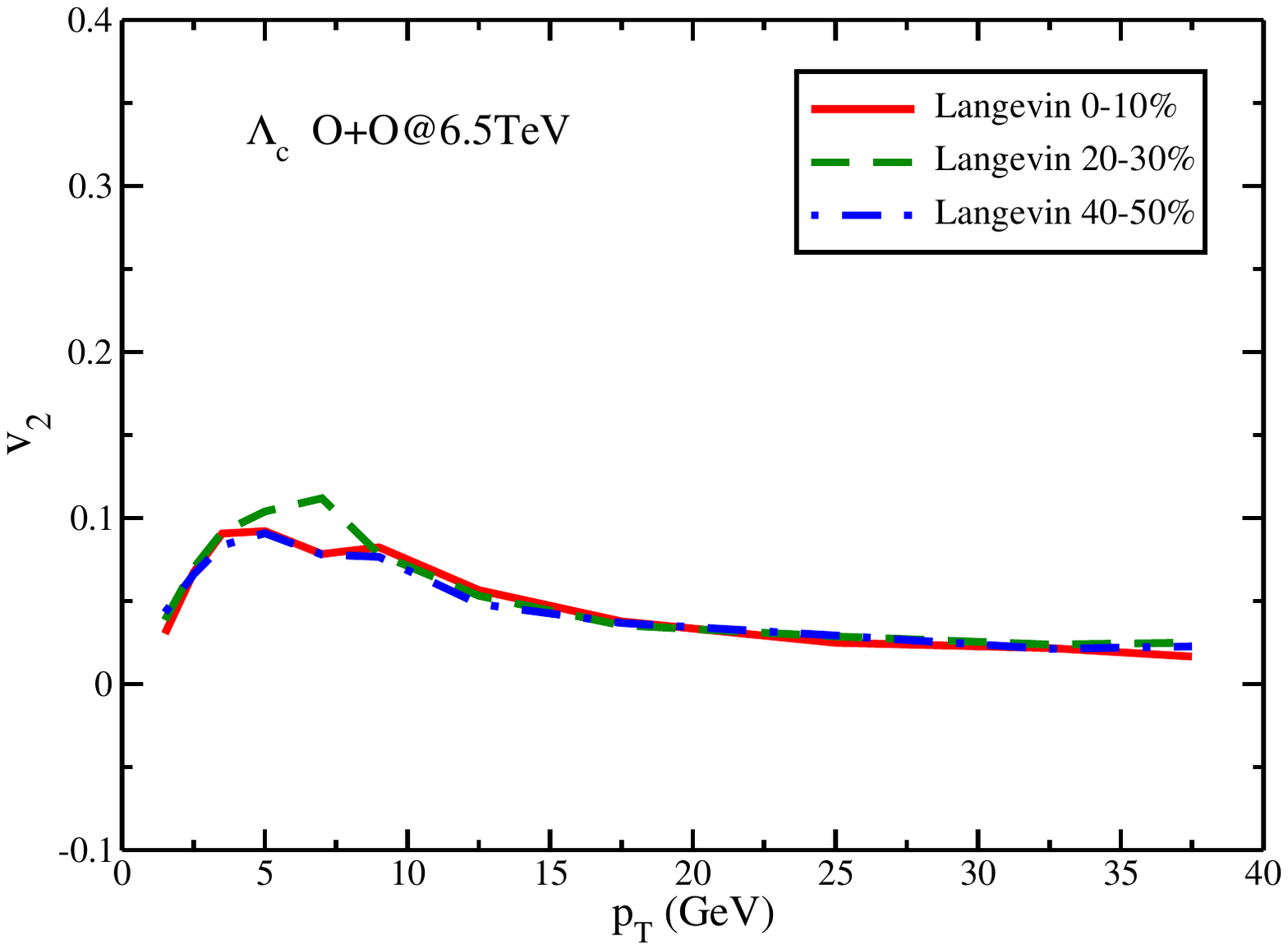}
\caption{The elliptic flow $v_2$ for $D^0$, $D_s$ and $\Lambda_c$ as a function of $p_T$ in central 0-10\%, 20-30\% and 40-50\% O+O collisions at $\sqrt{s_{NN}}=6.5$~TeV.}
\label{fig_v2_OO}
\end{figure}

\begin{figure}[tbh]
\includegraphics[width=0.89\linewidth]{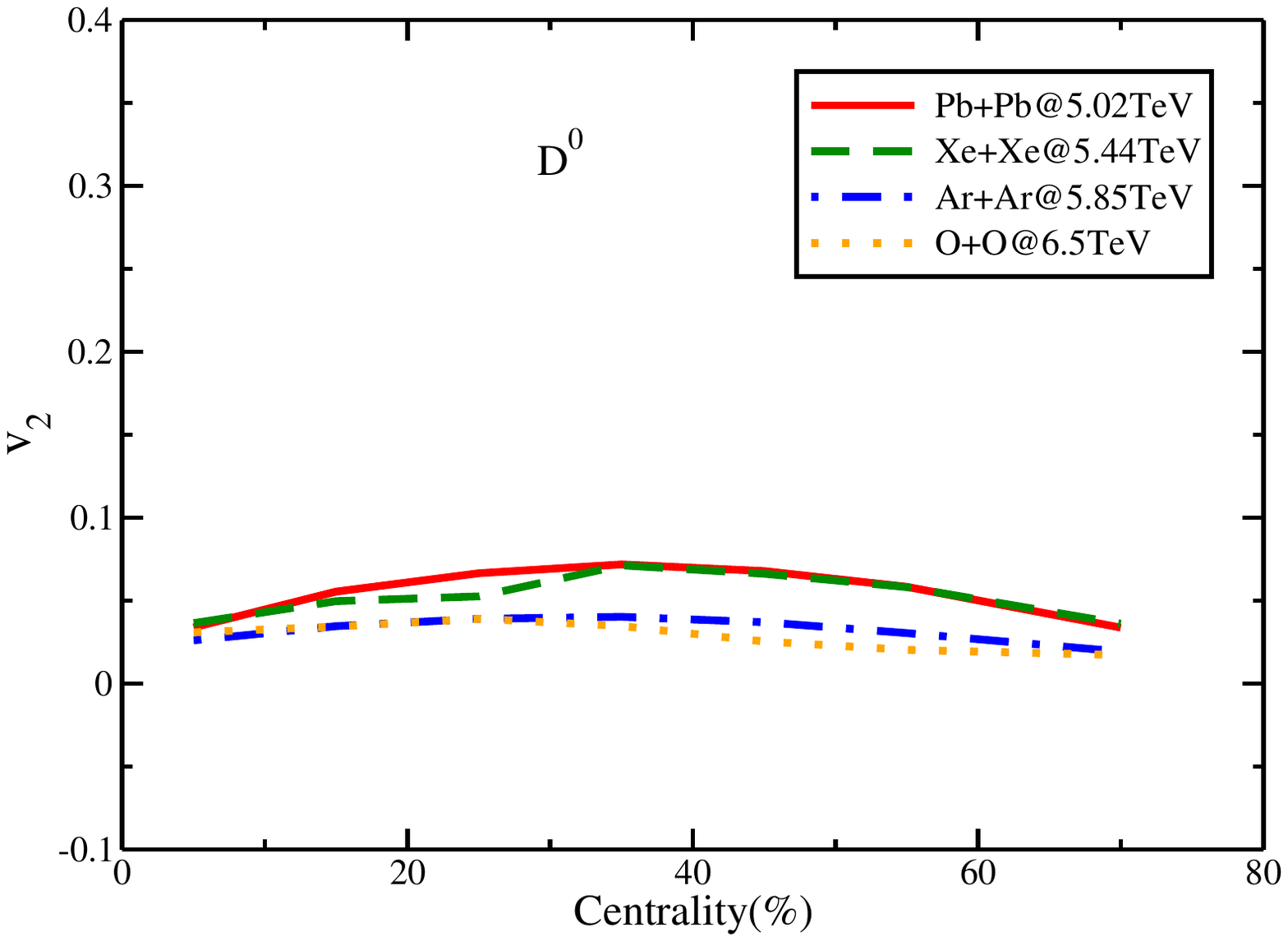}
\includegraphics[width=0.89\linewidth]{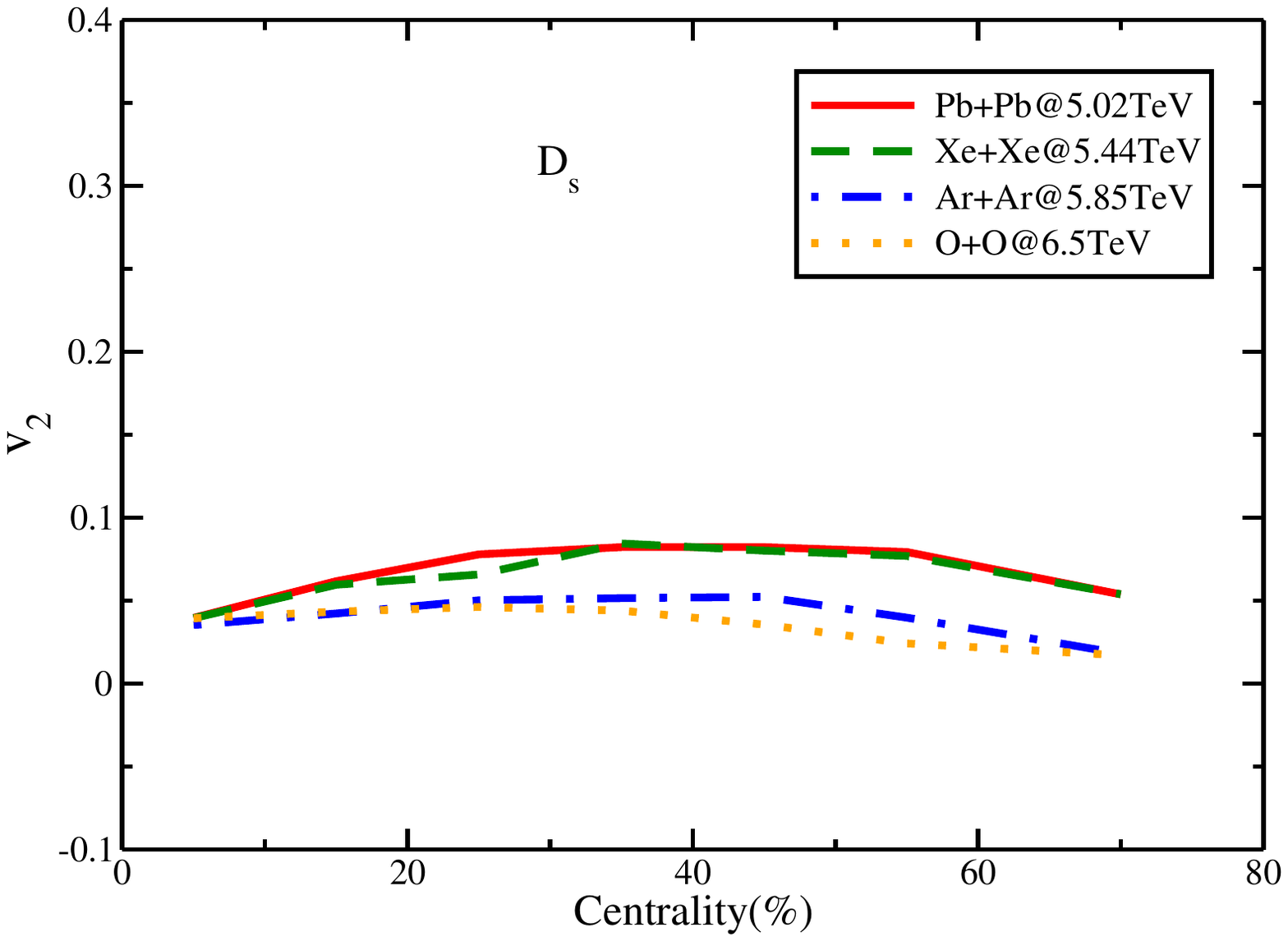}
\includegraphics[width=0.89\linewidth]{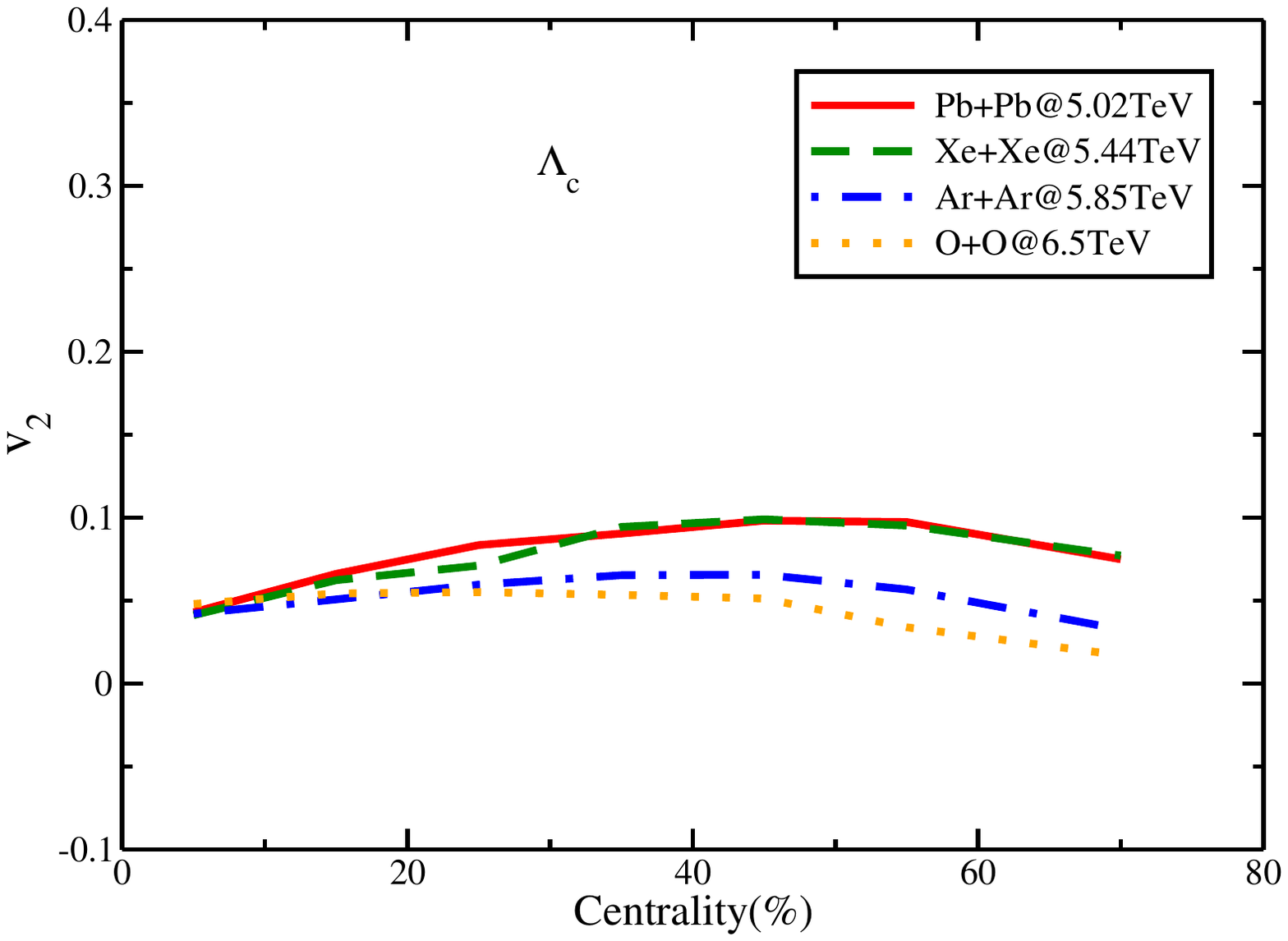}
\caption{The elliptic flow $v_2$ for $D^0$, $D_s$ and $\Lambda_c$ as a function of centrality in Pb+Pb, Xe+Xe, Ar+Ar and O+O collisions at the LHC energies.}
\label{fig_v2_centrality}
\end{figure}

\subsection{Charmed hadron flow}

In this work, the elliptic flow $v_2$ is calculated from the final state charm hadron momentum distribution $dN/d^2p_T dy$ as follows:
\begin{align}
\label{eq:v2}
v_2(p_{T}) = \langle \cos(2\phi)\rangle  = \left\langle \frac{p^{2}_{x} - p^{2}_{y}}{p^{2}_{x} + p^{2}_{y}}\right\rangle,
\end{align}
where $\langle \ldots \rangle$ represents the average over final state charmed hadrons generated in our simulations for each centrality class, and the $x$-$y$ axes define the transverse plane.
Note that in our study, we use the smooth hydrodynamic profiles for simulating heavy quarks interacting with the QGP medium, therefore the event plane of the ion collisions is defined by the $x$-$z$ axes.
Each smooth hydrodynamic profile is generated from an initial entropy density distribution by averaging over 5000 Trento events.
In order to capture the key effect of event-by-event fluctuations in the initial states on elliptic flow $v_2$ calculation, we rotate the participant plane of each Trento event to the event plane ($x$-$z$ plane) of our computational frame~\cite{Cao:2014fna,Cao:2017umt}.

Figure~\ref{fig_v2_PbPb} shows the elliptic flow $v_2$ of $D^0$, $D_s$ and $\Lambda_c$ as a function of $p_T$ in Pb+Pb collisions at $\sqrt{s_{NN}}=5.02$~TeV.
The results are shown for three different centralities (0-10\%, 10-30\% and 30-50\%) from our model calculations, and one centrality (30-50\%) from the ALICE data on $D^0$ and $D_s$ mesons~\cite{Vermunt:2019ecg,ALICE:2020iug,ALICE:2021kfc}.
Similarly, one can clearly observe the strong centrality dependence for the elliptic flow $v_2$. From most central (0-10\%) to mid-central (30-50\%) collisions, $v_2$ increases. This is mainly the effect of collision geometry: moving from central to mid-central and peripheral collisions, the elliptic eccentricity of the overlap region increases.
One would expect that when going to very peripheral collisions, even though the eccentricity continues to increases, $v_2$ will decrease due to the decreasing of the system size (see Fig.~\ref{fig_v2_centrality} later), since the system does not have enough time to fully develop the flow.

Figure~\ref{fig_v2_XeXe} shows our prediction for the elliptic flow $v_2$ of $D^0$, $D_s$ and $\Lambda_c$ as a function of $p_T$ in Xe+Xe collisions $\sqrt{s_{NN}}=5.44$~TeV for three centralities (0-10\%, 20-30\% and 40-50\%).
Compared to Pb+Pb results, the elliptic flow $v_2$ for charmed hadrons in Xe+Xe collisions are similar to those in Pb+Pb collisions.
This situation is very similar to the charmed hadron ratios, i.e., the system size effect is not very important between Pb+Pb and Xe+Xe collisions.
While the system size for Xe+Xe collisions is smaller than Pb+Pb collisions, the QGP produced in Xe+Xe collisions are still large enough to develop the radial flow as well as elliptic flow.

In Figs.~\ref{fig_v2_ArAr} and \ref{fig_v2_OO}, we show our prediction for the $D^0$, $D_s$ and $\Lambda_c$ elliptic flow $v_2$ as a function of $p_T$ in Ar+Ar collisions at $\sqrt{s_{NN}}=5.85$~TeV and in O+O collisions at $\sqrt{s_{NN}}=6.5$~TeV for three centralities (0-10\%, 10-20\% and 40-50\%).
Compared to Pb+Pb and Xe+Xe collisions, the centrality dependence for charmed hadron elliptic flow $v_2$ are much weaker in Ar+Ar and O+O collisions.
In particular, the elliptic flow $v_2$ in 40-50\% Ar+Ar collisions is slightly smaller than that in 20-30\% Ar+Ar collisions.
This means that for smaller Ar+Ar collisions, the system size effect starts to dominate from 40-50\% centrality.
For even smaller O+O collisions, one can see that the elliptic flow $v_2$ for the 0-10\%, 20-20\% and 40-50\% centralities are quite similar.
This means that the system size effect has already played the dominant role in 20-30\% centrality in O+O collisions.

To compare the charmed hadron elliptic flow among different collision systems more directly, Fig.~\ref{fig_ratio_centrality} shows the elliptic flow $v_2$ of $D^0$, $D_s$ and $\Lambda_c$ (integrated over $p_T$) as a function of collision centrality in Pb+Pb, Xe+Xe, Ar+Ar and O+O collisions.
One can clearly see that the values of elliptic flow $v_2$ are similar between Pb+Pb and Xe+Xe collisions.
As for Ar+Ar and O+O collisions, the elliptic flow $v_2$ values are much smaller compared to Pb+Pb and Xe+Xe collisions due to much smaller system sizes.
Another interesting feather is that the centrality dependence of elliptic flow $v_2$ in Ar+Ar and O+O collisions is much weaker than Pb+Pb and Xe+Xe collisions.
In particular, due to the competition between system size and collision geometry, the $v_2$ values in Ar+Ar and O+O collisions remain roughly constant in a wide rage of centrality (from 0-10\% centrality to 40-50\% centrality), and only have significant decrease in very peripheral collisions.


\section{Summary}

In this work, we have studied the charmed hadron ratios and elliptic flow in Pb+Pb, Xe+Xe, Ar+Ar and O+O collisions at the LHC energies.
The dynamical evolution of bulk QGP medium produced in heavy and light ion collisions is simulated using the (3+1)-dimensional CLVisc hydrodynamics model.
The evolution dynamics of heavy quarks inside the QGP is computed via a modified Langevin approach, which takes into account both collisional and radiative processes experienced by heavy quarks.
The hadronziation of heavy quarks is simulated with our hybrid coalescence-fragmentation model.

We first present the numerical results for the charmed hadron chemistry in terms of $D_s/D^0$ and $\Lambda_c/D^0$ ratios as a function of $p_T$ for different centralities in Pb+Pb collisions at $\sqrt{s_{NN}}=5.02$~TeV at the LHC.
Our model calculation shows a strong centrality dependence for charmed hadron ratios $D_s/D^0$ and $\Lambda_c/D^0$, which comes from the strangeness enhancement and parton coalescence effects.
In addition, the centrality dependence for $\Lambda_c/D^0$ is stronger than $D_s/D^0$, as a result of baryon-to-meson enhancement effect.
These are important signatures for the QGP formation in heavy-ion collision at RHIC and the LHC
We then present the results for charmed hadron chemistry in Xe+Xe, Ar+Ar and O+O collisions at the LHC energies.
It is interesting to observe that the $D_s/D^0$ and $\Lambda_c/D^0$ ratios in Xe+Xe collisions are similar to those in Pb+Pb collisions.
As for two light ion collisions (Ar+Ar and O+O), the charmed hadron ratios are also significantly smaller due to much smaller system sizes.
However, the strong centrality dependence still exists for the $D_s/D^0$ and $\Lambda_c/D^0$ ratios in Ar+Ar and O+O collisions, which can be regarded as an important signature for the formation of QGP, once confirmed by the experiments.

Later on, we present the elliptic flow $v_2$ as a function of $p_T$ for $D^0$, $D_s$ and $\Lambda_c$ in Pb+Pb, Xe+Xe, Ar+Ar and O+O collisions.
Similar to charmed hadron ratios, a strong centrality dependence is also observed for $D^0$, $D_s$ and $\Lambda_c$ elliptic flow $v_2$ in Pb+Pb and Xe+Xe collisions, as a result of the competing effects between system size and collision geometry.
However, for Ar+Ar and O+O collisions the centrality dependence of charmed hadron $v_2$ is much weaker.
In particular, the $v_2$ values for $D^0$, $D_s$ and $\Lambda_c$ remain roughly the same from 0-10\% to 40-50\% centralities, and only decreases significantly in very peripheral collisions.
This indicates that for light ion collisions (Ar+Ar and O+O), the system size effect has already played very important roles for elliptic flow in not-very-peripheral collisions.

In summary, we have presented a systematic study on charmed hadron chemistry and elliptic flow in both heavy and light ion collisions at the LHC energies.
Our study constitutes a significant reference for deciphering the evolution dynamics and hadronization mechanics of heavy quarks in hot and dense QCD matter.
It also sheds light on understanding the difference and similarity between large and small systems in relativistic nuclear collisions.


\section*{Acknowledgments}

This work is supported in part by the National Natural Science Foundation of China (NSFC) under Grant Nos. 12225503, 11890710, 11890711, 11935007, 12175122 and 2021-867. W.-J.~X. is supported in part by China Postdoctoral Science Foundation under Grant No. 2023M742099. Some of the calculations were performed in the Nuclear Science Computing Center at Central China Normal University (NSC$^3$), Wuhan, Hubei, China.

\bibliographystyle{plain}
\bibliographystyle{h-physrev5}
\bibliography{refs_GYQ}
\end{document}